\documentclass[preprint,12pt]{elsarticle}
\usepackage{amsmath,amssymb,amsfonts}
\usepackage{algorithmic}
\usepackage{graphicx}
\usepackage{subcaption}
\usepackage{framed}
\usepackage{textcomp}
\usepackage{hyperref}
\usepackage{xcolor}
\usepackage{tabularx}
\usepackage{listings}
\usepackage{array}
\usepackage{adjustbox}
\usepackage{float}
\usepackage{makecell}
\usepackage{multirow}
\usepackage{xltabular}
\usepackage{siunitx}
\usepackage{ragged2e}
\usepackage{wrapfig}
\usepackage{booktabs} 
\usepackage{svg}

\newcommand{\eg}{{\it e.g.}}

\def\BibTeX{{\rm B\kern-.05em{\sc i\kern-.025em b}\kern-.08em
    T\kern-.1667em\lower.7ex\hbox{E}\kern-.125emX}}
\begin{document} 

\begin{frontmatter}

\title{Improving MPI Error Detection and Repair \\ with Large Language Models and Bug References}





\author[1]{Scott Piersall}
\ead{spiersall@ucf.edu}

\author[1]{Yang Gao}
\ead{yang.gao@ucf.edu}

\author[1]{Shenyang Liu}
\ead{shenyang.liu@ucf.edu}

\author[1]{Liqiang Wang\corref{cor1}}
\ead{liqiang.wang@ucf.edu}
\cortext[cor1]{Corresponding author}

\affiliation[1]{organization={Dept. of Computer Science, University of Central Florida},
             city={Orlando},
             postcode={32816},
             state={FL},
             country={US}}
            
\begin{abstract}
Message Passing Interface (MPI) is a foundational technology in high-performance computing (HPC), widely used for large-scale simulations and distributed training (\eg, in machine learning frameworks such as PyTorch and TensorFlow). However, maintaining MPI programs remains challenging due to their complex interplay among processes and the intricacies of message passing and synchronization. With the advancement of large language models like ChatGPT, it is tempting to adopt such technology for automated error detection and repair. Yet, our studies reveal that directly applying  large language models (LLMs) yields suboptimal results, largely because these models lack essential knowledge about correct and incorrect usage, particularly the bugs found in MPI programs. In this paper, we design a bug detection and repair technique alongside Few-Shot Learning (FSL), Chain-of-Thought (CoT) reasoning, and Retrieval Augmented Generation (RAG) techniques in LLMs to enhance large language model’s ability to detect and repair errors. Surprisingly, such enhancements lead to a significant improvement, from 44\% to 77\%, in error detection accuracy compared to baseline methods that use ChatGPT directly. Additionally, our experiments demonstrate our bug referencing technique generalizes well to other large language models.
\end{abstract}

\begin{keyword}
MPI Programs, Large Language Models, Error Detection and Repair
\end{keyword}

\end{frontmatter}

\section{Introduction}

High-performance computing (HPC) is crucial in numerous fields, including information assurance, healthcare, computational sciences, and machine learning. The Message Passing Interface (MPI) is a widely used programming library in high-performance computing. It follows the Single Program, Multiple Data (SPMD) processing model, where the same program is executed across multiple processes, each with its own private address space. MPI provides a set of library routines to facilitate communication and synchronization through message exchanges between processes. Parallel programs developed using MPI can introduce unique, complex, and difficult-to-identify errors that have no direct counterparts in serial programming, challenging traditional defect detection techniques. 

MPI interface contains complex features including non-blocking communication, non-deterministic execution, and collective operations which require precise sequential ordering. Fault isolation is especially challenging in the presence of defects when MPI program behavior differs among different executions, a result of the non-deterministic behavior of MPI.  Additionally, these challenges are amplified when varying the number of processing nodes among program executions produces different results as a side-effect when defects are present.

Traditional software defect detection techniques, such as static analysis, dynamic analysis, symbolic execution, and concolic testing, have been employed to address the challenges of identifying defects in MPI programs. However, these methods only address a limited subset of the diverse range of MPI defects. Additionally, they face significant limitations, including high false positive rates, scalability challenges in analyzing large programs, and incomplete path coverage for complex software. These drawbacks render such approaches impractical for comprehensive defect detection in real-world MPI applications.

Large language models (LLMs) demonstrate significant potential in detecting and repairing software defects. Research into their effectiveness spans various computing domains. LLMs identify software defects through pattern recognition, contextual understanding, and by simulating static code analysis tools. Trained on vast datasets containing millions of lines of code, they can detect patterns associated with defects or syntactic errors. Additionally, LLMs analyze code flow to identify issues and apply software engineering best practices to detect deviations that may result in errors.

Our work proposes a novel approach to improve model out-of-the-box accuracy in detecting and repairing defects in MPI programs. 
Notably, our objective is not to compare the performance of different LLMs; instead, we propose a method applicable to any LLM that employs CoT reasoning, Few-shot learning (FSL), and Retrieval-Augmented Generation (RAG), and compare how these methods improve MPI error detection and repair on several typical LLMs such as ChatGPT, Llama2, QWen2.5-coder, and Code Llama.

We conducted a series of experimental trials to measure the improvement our defect detection approach provides. We compared our results with the baseline performance of ChatGPT without our approach. Our experiments demonstrated that our bug reference approach enhances defect detection accuracy by 75\% compared to the direct use of ChatGPT. Using this improvement in detection accuracy, we evaluated ChatGPT's capability to repair defective MPI software code. Finally, we conducted experiments using our approach on other LLMs to observe the applicability of our approach on other LLMs. 

Our research focuses on known defective MPI programs obtained from a publicly available dataset. For our study, we selected a subset of these MPI programs where traditional analysis techniques demonstrated high rate of false negatives, indicating difficulty in identifying defects \cite{MPI_BI}. We recognize an urgent need to enhance the accuracy of LLMs in detecting and repairing defects in parallel programs. To the best of our knowledge, this work represents the first effort to employ targeted strategies for improving LLM performance in identifying and repairing software defects in MPI programs, specifically those that traditional methods have struggled to address \cite{MPI_BI}.

\section{Related Work}


MPI is a foundational technology for high-performance computing (HPC), enabling scalable parallelism in large-scale scientific simulations and distributed machine learning. While MPI provides efficient distributed-memory communication, its non-determinism, wildcard receives, and concurrency semantics introduce significant complexity, making defect detection and debugging challenging.

\subsection{Classic Error Detection and Repair in MPI Programs}

A substantial body of work has addressed MPI defect detection, yet long-standing challenges persist due to non-deterministic communication and complex synchronization behavior.


\textbf{Static analysis} inspects source code to detect issues such as type mismatches, incorrect buffer usage, and logical errors. Tools such as MPI-CHECK \cite{Droste_2015_MPI_CHECKER_STATIC} improve path sensitivity for MPI primitives but remain limited in handling wildcard communication and dynamic memory behavior, often producing high false-positive rates and missing runtime-dependent defects \cite{MPI_BI}. Hybrid approaches have been proposed to mitigate these limitations. For example, Ma \textit{et al.} reformulate MPI thread-safety violations as race conditions by injecting narrowly scoped wrapper-based instrumentation around specific MPI calls, achieving high accuracy and low overhead, though restricted to a single defect class \cite{ma2015detecting}.


\textbf{Dynamic analysis} detects defects by monitoring program execution, enabling identification of deadlocks, synchronization errors, and resource leaks \cite{Vakkalanka_2008_ISP_ModelChecking, Hilbrich_2012_MUST_DeadlockDetection}. However, scalability remains a key limitation, as instrumentation overhead grows with process count and analysis is limited to exercised execution paths \cite{Li_2018_COMPI_Concolic, Hilbrich_2012_GTI_Parallel}. Tools such as MUST, MPI-CHECK, and ISP integrate dynamic analysis and model checking to improve coverage, though at significant runtime cost \cite{Droste_2015_MPI_CHECKER_STATIC, Vakkalanka_2008_ISP_ModelChecking, Hilbrich_2012_MUST_DeadlockDetection, Li_2019_Efficient_Concolic}.



\textbf{Symbolic execution} and \textbf{concolic testing} aim to explore execution paths exhaustively by treating inputs symbolically, offering stronger theoretical coverage guarantees \cite{siegel2011automatic, Chen_2020_MPI_SV, 10479444}. In practice, these approaches suffer from state-space explosion and are further complicated by MPI non-determinism and large input domains, limiting applicability to small or simplified programs \cite{Li_2018_COMPI_Concolic, Li_2019_Efficient_Concolic}.

While these classical techniques have proven effective for specific defect classes and limited program scopes, they continue to face fundamental challenges related to scalability, false positives, and incomplete coverage, constraining their practical use for large, real-world MPI applications \cite{Vakkalanka_2008_ISP_ModelChecking, Li_2019_Efficient_Concolic, siegel2011automatic, Chen_2020_MPI_SV, 10479444, 10025545}.

\subsection{Modern Software Defect Detection and Repair with LLMs}


Recent advances in large language models (LLMs), such as GPT-based systems, have generated significant interest for automated software defect detection, debugging, testing, and repair. These models demonstrate strong capabilities in learning code syntax and semantics and are increasingly applied across multiple software engineering tasks.


\textbf{Debugging using LLMs} focuses primarily on fault localization. Approaches such as SoapFL \cite{SOAPFL_LLM_LocalFaultIsolation} and FlexFL \cite{FLEXFL_LLM_LocalFaultIsolation} leverage LLMs to identify defective functions, with FlexFL improving generalization by removing the requirement for failed unit tests. However, these techniques remain constrained by limited context windows and often depend on auxiliary artifacts, such as test suites, which are frequently incomplete in legacy code bases \cite{LLMS_SE_SURVEYANDOPENPROBLEMS, STUDY_FAULT_LOCALIZATION}.

\textbf{Defect detection with static analysis and LLMs} combines traditional analyzers with LLM reasoning to reduce false positives and false negatives \cite{LI_ENHANCINGSTATICANALYSIS_WITHLLM, ACM_LLM_STATIC_ANALYSIS, WI_ADVSCANNER_STATIC_LLM}. LLift \cite{LI_ENHANCINGSTATICANALYSIS_WITHLLM} integrates static analysis outputs with LLM prompts to detect uninitialized-variable defects in the Linux kernel, but its scope is limited to a single defect type, requires manual prompt construction, and reports a high false-positive rate. AdvScanner \cite{WI_ADVSCANNER_STATIC_LLM} improves detection effectiveness by combining multiple analyzers with LLM reasoning but remains narrowly focused on reentrancy defects and function-level analysis, limiting applicability to MPI programs.



\textbf{Software testing} has also benefited from LLMs, particularly for automated test generation. Systems such as TECO \cite{LLM_SOFTWARETESTING_TECO} and CAT-LM \cite{LLM_SOFTWARE_TESTING_CAT-LM} leverage semantic representations to generate higher-quality tests, demonstrating notable accuracy improvements over syntax-based baselines. Nevertheless, test generation presents unique challenges that often require specialized training or fine-tuning \cite{LLMS_SE_SURVEYANDOPENPROBLEMS, SoftwareTesting_with_LLMS_Wang_10440574}.



\textbf{Software defect repair using LLMs} integrates LLM reasoning with automatic program repair (APR). Models such as Codex and CodeBERT-based systems have demonstrated the ability to generate semantically meaningful patches \cite{LAJKO_APR_2024, APR_SequenceR}. Despite these advances, LLM-based repair remains sensitive to prompt design, computationally expensive, and often requires domain-specific knowledge to produce reliable results \cite{APR_ZHAO_2024, CodeBERT_FENG_2020}.



LLMs trained on large code corpora have shown promise in detecting defects and suggesting refinements by jointly reasoning over syntax and context \cite{GUO_ChatGPT_CodeRefinement, Mashhadi_2021_CodeBERT_RepairJava, Xiong_2024_CodeBERT_FineTuning}. Prompt engineering has further improved performance across defect detection and repair tasks \cite{cao2023studypromptdesignadvantages, zhang2024promptenhancedsoftwarevulnerabilitydetection}. However, MPI-specific challenges—such as deadlocks, race conditions, and collective mismatches—remain underexplored, suggesting that MPI-aware fine-tuning or domain-informed prompting may be necessary to fully realize LLM potential in this domain.



Commercial LLMs, including OpenAI Codex, Anthropic Claude Code, and Google Gemini Code Assist, alongside open-weight models such as Code Llama and Qwen2.5-Coder, are increasingly adopted for software engineering workflows. While these models improve developer productivity, they remain imperfect defect detectors, often over-trusting plausible code, missing subtle semantic or concurrency-related faults, and exhibiting sensitivity to incomplete specifications or noisy prompts, resulting in both false negatives and false positives.

\section{Study Design}

\subsection{LLM Configuration}
Our initial experiments employed the ChatGPT GPT-3.5 Turbo model, integrated via the OpenAI v1.54.3 API with ChromaDB v0.5.18 and LangChain v0.3.7. To mitigate the model’s non-deterministic behavior, each query was executed five times using independent API conversations. To assess the generality of our approach, we replicated the same experimental protocol using open-source LLMs including Llama2, Qwen2.5-Coder, and Code LlaMA, all accessed through the Ollama-0.6.1 Python library and selected for their availability and parameter sizes comparable to GPT-3.5 Turbo (approximately 13 billion parameters).




\subsection{Dataset Pre-processing}
The MPI Bugs Initiative dataset \cite{MPI_BI} contains defect-descriptive comments in defective program samples, which are uncommon in real-world MPI software. We therefore removed all source code comments to better reflect realistic MPI code and prevent defect-specific annotations from influencing the analysis.

\section{Bug Detection of MPI Programs}

In this section, we established a baseline by evaluating ChatGPT's performance in detecting MPI-specific bugs, without specialized guidance, through the use of Zero-Shot prompt. The results indicated that ChatGPT, in its default state, has difficulty identifying complex parallel programming errors, underscoring the necessity for more targeted methodologies. Subsequently, we demonstrated how incorporating bug examples and domain-specific context significantly improves ChatGPT's performance. To achieve this enhancement, we integrated these examples with advanced techniques, including Few-Shot Learning, Chain-of-Thought (CoT) reasoning, and Retrieval-Augmented Generation (RAG). Combining these techniques leads to a notable increase in detection accuracy for complex MPI-specific defects, particularly those that lack direct analogs in serial programming.

\subsection{Core Definitions}
In software defect analysis, the confusion matrix terms are defined relative to whether a real defect exists in software code and whether the analysis method reports a defect \cite{MPI_BI}  \cite{Baldoni2018_SymbolicSurvey}. We utilize the confusion matrix terms in the context of software defect analysis. Table \ref{tab:defect_confusion} depicts the terms, definitions, and interpretations used in our trials, study, and subsequent analysis. 

\begin{table}[ht]
\centering
\caption{Core Definitions in Software Defect Detection}
\label{tab:defect_confusion}
\begin{adjustbox}{max width=0.85\linewidth}
    
\begin{tabular}{l|m{4.3cm}|m{3.3cm}}

\textbf{Term} & \textbf{Definition} & \textbf{Interpretation} \\
\hline
True Positive (TP) &
The analysis reports a defect and the defect is truly present in the code &
Correctly detected software defect \\
\hline
True Negative (TN) &
The analysis reports no defect and the code is truly defect-free &
Correctly ignored non-defective code \\
\hline
False Positive (FP) &
The analysis reports a defect when no defect exists in the code &
Spurious warning or false alarm \\
\hline
False Negative (FN) &
The analysis reports no defect when a defect is actually present &
Missed software defect \\

\end{tabular}
\end{adjustbox}
\end{table}

\subsection{Baseline with Zero-Shot}
\label{sec:basezeros}

We first established a zero-shot baseline to evaluate ChatGPT’s ability to detect MPI errors without domain-specific guidance. Using the basic prompt shown below, we conducted five independent trials over all 241 dataset samples and reported the average accuracy, precision, recall, and F1 score. This baseline assesses the model’s inherent capability to identify MPI-specific defects.

\begin{framed}
\textbf{\underline{Prompt:}} Does the following MPI program \textit{[CODE]} contain bugs?
\end{framed}

To establish the initial baseline, we replaced the placeholder \textit{[CODE]} in this basic prompt with each of the 241 MPI programs in our dataset. This process was repeated for  five separate trials, and the results from these trials are summarized in the first row of Table \ref{tab:ablation_details_summary}.



The zero-shot baseline produced poor results, with ChatGPT exhibiting a false negative rate of 46.88\% (113 of 241 samples on average; Table \ref{tab:ablation_details_summary}), rendering it unsuitable for practical MPI defect detection. In particular, the model failed to identify straightforward defects such as resource leaks caused by unreleased dynamically allocated resources as illustrated in Listing \ref{listing:ResourceLeakListing}. These findings suggest that GhatGPT's training data likely contained few defective annotated MPI programs, motivating our use of learning enhancement techniques. We specifically utilize Chain-of-Thought, Few-Shot Learning, and Retrieval-Augmented Generation learning enhancement techniques in subsequent experiments to improve MPI defect detection performance.

\begin{framed}
\textbf{Result}: ChatGPT exhibits limited out-of-the-box accuracy in identifying software defects in MPI programs, and our analysis, coupled with the high false-negative rate observed in the baseline trial indicates that domain-specific guidance, particularly detailed information on correct MPI library function ordering and usage, is necessary to substantially improve performance.

\end{framed}
\label{sec:rq1}

\subsection{Enhancing defect detection via explicit bug referencing}
We integrate explicit bug references into few-shot learning (FSL), chain-of-thought (CoT) reasoning, and Retrieval-Augmented Generation (RAG), together with expert guidance, to evaluate their impact on ChatGPT’s accuracy in detecting MPI software defects. To support these methods, we construct a domain-informed MPI bug repository guided by the prevalence of common defect classes reported in prior works, including deadlocks from mismatched communications, resource leaks and incorrect results from improper memory management, and concurrency and synchronization errors due to incorrect MPI API usage \cite{Chen_2020_MPI_SV,Baldoni2018_SymbolicSurvey,Droste_2015_MPI_CHECKER_STATIC,Hilbrich_2012_GTI_Parallel,MPI_BI,HUang_2020_Predictive_Deadlock}. The repository targets representative defect categories—resource leaks, mismatched collective and point-to-point communications, invalid or mismatched parameters, and incorrect synchronization—to systematically assess which combinations of examples and expert-informed prompting yield the greatest improvement in defect detection accuracy.


\subsubsection{Incorporating Bug References via Few-Shot Prompts}

To reduce the high false negative rate observed in our baseline experiments, we incorporated curated bug references into Few-Shot prompts, a technique that guides large language models using a small number of illustrative input–output examples without requiring model fine-tuning. Specifically, we modified the Zero-Shot prompt from Section~\ref{sec:basezeros} to include demonstrations of defective MPI programs, each annotated with explanations of the defect type and precise location (e.g., line number). These demonstrations covered a range of MPI defect classes, including resource leaks, incorrect call ordering, invalid parameters, deadlocks, and point-to-point and collective communication errors. This design enabled a controlled evaluation of how explicit defect examples influence model performance, with the resulting Few-Shot prompt presented below.

\begin{framed}
\textbf{\underline{Few-Shot Prompt:}} 
Here are some examples of MPI programs. 

\textit{[EXAMPLE1]} \textit{[EXPLANATION1]}

\textit{[EXAMPLE2]} \textit{[EXPLANATION2]}

...

\textit{[EXAMPLE9]} \textit{[EXPLANATION9]}

Does the following MPI program \textit{[CODE]} contain bugs? Respond with your VERDICT
\end{framed}

\begin{wrapfigure}{r}{0.38\textwidth}
\centering
   \includegraphics[width=0.35\textwidth]{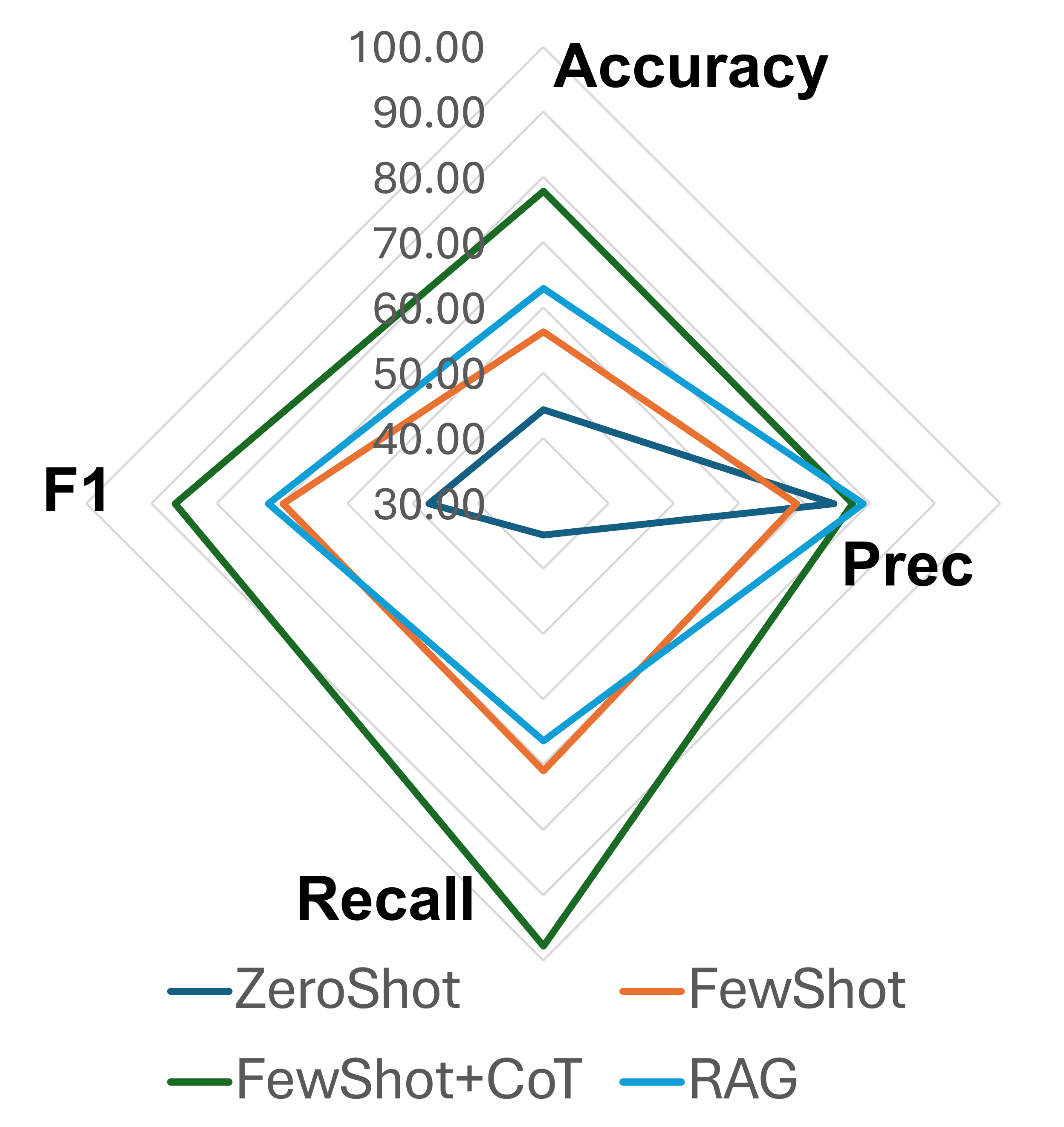}
    \caption{Comparison of Zero-Shot, Few-Shot, Few-Shot+Chain-of-Thought (CoT), and  Few-Shot+CoT+RAG prompting techniques. The inclusion of Few-Shot and CoT reasoning significantly enhances performance across all metrics.}
    \label{fig:gptzfc}
\end{wrapfigure}

Our Few-Shot \textit{[EXAMPLE(S)]} of defective MPI programs included detailed \textit{[EXPLANATION(S)]}, including a description of the defect, the line number of the defect, and suggested repair steps. An example bug reference used in our Few-Shot prompts is included in Listing 2.

We conducted a series of Few-Shot trials in which \textit{[CODE]} was systematically replaced with each of the 241 pre-processed programs from our testing dataset. A total of five trials were performed. 

\begin{table}[ht]
    \centering
    \caption{Quantitative evaluation of LLM prompting strategies (Zero-Shot, Few-Shot, Few-Shot+CoT and Few-Shot+CoT+Retrieval-Augmented Generation (RAG) with varying retrieval coverage), reporting classification performance across standard metrics and highlighting the superior performance of Few-Shot+CoT in terms of accuracy, recall, and F1-score. CoT denotes Chain-of-Thought reasoning; TP, TN, FP, and FN denote true positives, true negatives, false positives, and false negatives, respectively. A complete listing of all trials is provided in \ref{tab:appendix_chatgpttrials}.}
    \begin{adjustbox}{max width=\linewidth}
    \begin{tabular}{>{\raggedright\arraybackslash}m{3.4cm}|cccc|cccc} 
         \textbf{Prompting Tech}&\textbf{TP}&\textbf{TN}&\textbf{FP}&\textbf{FN}&\textbf{Acc(\%)}&\textbf{Prec(\%)}&  \textbf{Rec(\%)} & \textbf{F1(\%)} \\
        \Xhline{1pt}
        Zero-Shot & 60.6 & \textbf{46.4} & \textbf{20.6} &113.4 & 44.39 & 74.60&34.82 &47.46\\
         \hline 
        Few-Shot & 121.8 & 14.0 & 55.2 & 50.0 & 56.34 & 68.82 & 70.91 &69.83   \\
         \hline 
        \textbf{Few-Shot+CoT} & \textbf{170.2} & 17.4 & 49.6 & \textbf{3.8} & \textbf{77.84} & 77.43 & \textbf{97.81} & \textbf{86.43}  \\
         \hline 
        Few-Shot+CoT +RAG\_100\% & 115.8 &36.0 & 30.6 &58.6 &62.98 &\textbf{79.07} & 66.39 & 72.11 \\
         \hline 
        Few-Shot+CoT +RAG\_75\% &  103.6 &  37.4 &  29.6 &  70.4 &  58.50 &  77.77 &  59.54 &  67.39 \\
         \hline 
        Few-Shot+CoT +RAG\_50\%  &  101.2 &  36.4 &  30.6 &  72.8 &  57.09 &  76.77 &  58.16 &  66.17  \\
         \hline 
        Few-Shot+CoT +RAG\_25\%  &  105.6 &  36.6 &  30.4 &  68.4 &  59.00 &  77.70 &  60.68 &  68.05  \\
         \hline 
        Few-Shot+CoT +RAG\_0\% &  101 &  36.8 &  30.2 &  73 &  57.17 &  77.08 &  58.04 &  66.13  \\

    \end{tabular}
    \end{adjustbox}
    \medskip
    \label{tab:ablation_details_summary}
\end{table}

A comparison of the Few-Shot results in Table~\ref{tab:ablation_details_summary} with our Zero-Shot baseline reveals 11.95\% improvement in accuracy (from 44.39\% to 56.34\%). This gain is directly attributable to the inclusion of bug reference examples in the Few-Shot prompts, which provide concrete demonstrations of defective MPI code along with expert explanations.

Figure~\ref{fig:gptzfc} further visualizes this improvement by comparing classification performance in terms of accuracy, precision, recall, and F1-score under Zero-Shot and Few-Shot prompting. In the figure, the blue curve corresponds to the Zero-Shot configuration, which relies solely on task instructions and exhibits consistently weaker performance across all four metrics. In contrast, the orange curve represents the Few-Shot configuration augmented with bug reference examples, showing a clear and sustained improvement in accuracy, recall, and F1-score throughout the experimental trials. The pronounced separation between the two curves highlights the effectiveness of incorporating domain-specific defective code examples and explanations, enabling the model to better generalize to unseen MPI programs and achieve more reliable and balanced defect detection than instruction-only Zero-Shot prompting. Building on these findings, we expanded our experiments to investigate additional learning enhancement techniques.


\subsubsection{Incorporating bug references with Chain-of-Thought}
CoT reasoning has been shown to significantly improve the ability of LLMs, such as ChatGPT, to perform complex reasoning \cite{wei2022chain} tasks. Given the complex nature and non-deterministic behavior of MPI programs we conducted experimental trials using our Few-Shot prompt combined with CoT reasoning prompts. 

We developed nine CoT prompts by formulating straightforward descriptions of common MPI programming errors as a sequence of step-by-step tests. These nine CoT prompts included step-by-step guidance for: {\it MPI point-to-point communications, MPI collective communications, matching message sizes in MPI send and receive calls,  resource leaks resulting from unreleased dynamically-allocated MPI resources, MPI message races, mismatched MPI communication parameter tags, missing starts, MPI scatter with no gather}, and {\it missing wait operations}.

Our CoT prompts were then used to elicit ChatGPT's assessment of each test regarding the MPI code under evaluation. Each MPI program in our dataset was analyzed using all nine Chain-of-Thought prompts. A representative example of our Chain-of-Thought reasoning, integrated into our Few-Shot approach, is presented below.

\begin{framed}
\textbf{\underline{Few-Shot Bug References + Chain of Thought:}} 
Here are some examples of MPI programs.

\textit{[EXAMPLE1]} \textit{[EXPLANATION1]}

\textit{[EXAMPLE2]} \textit{[EXPLANATION2]}

...

\textit{[EXAMPLE9]} \textit{[EXPLANATION9]}

Does the following MPI program \textit{[CODE]} contain bugs? Step-by-step, check the code to see if:

1. Each point-to-point or collective communication is correct

2. The message sizes in the send and receive calls match

3. There are no message races or mismatched tags

4. There are no resource leaks

5. There are no missing starts or waits.

For each step, review the code in 100 words. Finally, at the end of your response, for each item, respond with your VERDICT.
For each item with definite errors, you must end your response with ``MAJOR ERRORS DETECTED", else you must end your response exactly with ``NO MAJOR ERRORS DETECTED".
\end{framed}

We conducted experimental trials combining our Few-Shot Bug Reference prompt with Chain-of-Thought reasoning. Each of the 241 pre-processed programs was evaluated across five independent trials. The results of these trials, using the Few-Shot prompt with Chain-of-Thought reasoning, are summarized in Table \ref{tab:ablation_details_summary}.

\begin{figure*}[ht]
    \centering
    \includegraphics[width=0.6\textwidth]{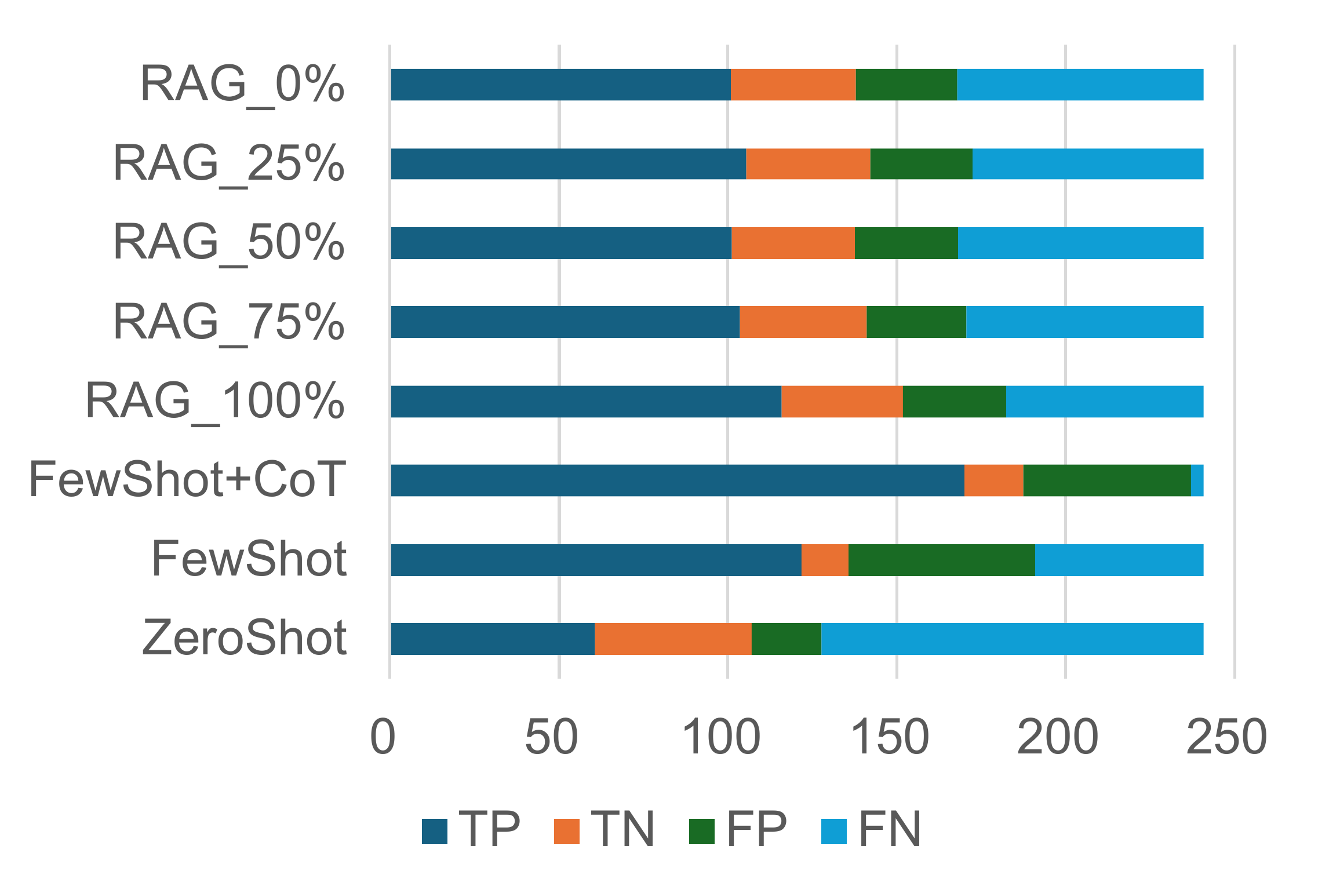}
    \caption{Detailed Performance Metrics Across Experimental ChatGPT Trials: Comparison of True Positives (TP), True Negatives (TN), False Positives (FP), and False Negatives (FN) across Zero-Shot, Few-Shot, CoT, and RAG experimental setups, providing an in-depth insight into the strengths and weaknesses of each method in identifying MPI program defects.}
    \label{fig:gptbar}
\end{figure*}

As shown in Table \ref{tab:ablation_details_summary} and depicted in Figures \ref{fig:gptzfc} and \ref{fig:gptbar} our experimental results from our Few-Shot with CoT reasoning provide significant improvements in identifying software defects in our MPI program dataset. Additionally, our average false negative rate across all five trials is 1.5 percent, indicating a high sensitivity for our approach.


\begin{framed}
\textbf{Results}: The addition of handcrafted MPI programs with defects, coupled with detailed explanations regarding the nature and location of the defects, provided an increase of 33 percentage points in ChatGPT's accuracy in identifying software defects over our out-of-the-box ChatGPT Zero Shot baseline.
\end{framed}
\label{sec:cot}



\begin{table}[ht]
    \centering
    \caption{Comparison of Variances $\sigma^2$ Among Experiments. Notably, our integrated approach produces the lowest variance for True Positives, True Negatives, False Positives, and False Negatives. CoT=Chain Of Thought Reasoning, TP=True Positives, TN=True Negatives, FP=False Positives, FN=False Negatives. A complete listing of all trials is depicted in \ref{tab:appendix_chatgpttrials}}
    \begin{adjustbox}{max width=0.76\linewidth}
    \begin{tabular}{l|cccc}
         \textbf{Prompting Tech}&\textbf{TP $\sigma^2$}&\textbf{TN $\sigma^2$}&\textbf{FP $\sigma^2$}&\textbf{FN $\sigma^2$} \\          
        \Xhline{1pt}   
        Zero-Shot & 17.84 & 9.04 & 9.04 & 17.84 \\
        \hline 
        Few-Shot & 3.76 & 4.8 & 8.96 & 6.4 \\
        \hline 
        \textbf{Few-Shot+CoT} & \textbf{3.76} & \textbf{2.24} & \textbf{2.24} & \textbf{3.76}  \\
        \hline 
        Few-Shot+CoT+RAG\_100\% & 62.56 &4.4 & 3.44 &60.64 \\
        \hline 
        Few-Shot+CoT+RAG\_75\% & 33.04 & 3.44 &  3.44 &  33.04\\
        \hline 
        Few-Shot+CoT+RAG\_50\%  &  12.56 &  5.84 &  5.84 &  12.56\\
        \hline 
        Few-Shot+CoT+RAG\_25\%  &  52.64 &  20.3 &  20.3 &  65.8\\
        \hline 
        Few-Shot+CoT+RAG\_0\% & 62.0 &  45.76 &  45.76  &  61.9\\
        
    \end{tabular}
    \end{adjustbox}
    \medskip
    \label{tab:chatgpt_variance_summary}
\end{table}

\begin{table}[ht]
    \centering
    \footnotesize
    \setlength{\tabcolsep}{4pt}
    \caption{Performance metrics of Non-LLM Approaches.}
    \begin{tabular}{l|cccc|cccc}  
         \textbf{Tool} &\textbf{TP}&\textbf{TN}&\textbf{FP}&\textbf{FN}&\textbf{Acc(\%)}&  \textbf{Prec(\%)}&  \textbf{Rec(\%)}& \textbf{F1(\%)}\\ 
        \Xhline{1pt}
           CIVL\cite{CIVL_LUO_2017}&61&18&61&101& 32.78& 37.65& 50.00& 42.96\\ \hline 
           ISP\cite{Vakkalanka_2008_ISP_ModelChecking}&116&17&94&14& 55.19& 89.23& 55.24& 68.24\\ \hline 
           MPISV\cite{Chen_2020_MPI_SV}&68&15&64&94& 34.44& 41.98& 51.52& 46.26\\ \hline 
           PARCOACH\cite{PARCOACH2020}&140&10&79&12& 62.24& 92.11& 63.93& 75.47\\ 
    \end{tabular}
    \label{tab:nonllmtools}
\end{table}

We calculated the variance ($\sigma^2$) of ChatGPT's results among each study across five trials and recorded them in Table \ref{tab:chatgpt_variance_summary}. A comparison of the variances demonstrates an additional noteworthy observation: our bug referencing approach has the lowest variance in True Positives, True Negatives, False Positives, and False Negatives. The lower variances delivered by our bug referencing technique demonstrate a higher reliability, stability, and less non-determinism in our defect analysis. Additionally, these lower variances in our bug referencing technique demonstrate improved and more consistent model reasoning.

\subsubsection{The Alternative - RAG with Good and Defective MPI Examples}

Retrieval Augmented Generation (RAG) augments the self-contained knowledge of large language models by dynamically incorporating domain-specific information at inference time, a capability shown to reduce hallucinations and improve accuracy in NLP tasks \cite{Hamed_2022_RAG, Salemi_2024_RAG_Evaluation}. Prior work demonstrates that RAG can enhance response quality by supplementing model knowledge with external data not present in the original training corpus, thereby improving accuracy and relevance in specialized domains without the cost of fine-tuning or retraining \cite{Jacobs_2024_RAG, Salemi_2024_RAG_Evaluation}. Motivated by these findings, we applied RAG to MPI defect detection by constructing retrieval databases composed of known correct and defective MPI source code \cite{Jacobs_2024_RAG} and evaluating five separate databases with varying ratios of good to buggy programs to assess how retrieval corpus composition influences defect detection accuracy.

\begin{table}[ht]
    \centering
    \caption{Nomenclature and description of retrieval databases used in RAG studies.}
    \begin{adjustbox}{max width=\linewidth}
    \begin{tabular}{l|c}  
         \textbf{Nomenclature}&\textbf{Description} \\   \Xhline{1pt}            
         RAG\_100\%&Corpus containing only known good MPI source code\\ 
         \hline 
         \multirow{2}{*}{RAG\_75\%}&Corpus containing 75\% defect free MPI source code and \\ 
          & 25\% MPI source code known to contain at least one defect \\
         \hline 
         \multirow{2}{*}{RAG\_50\%}&Corpus containing 50\% defect-free MPI source code and \\ 
          & 50\% MPI  source code known to contain at least one defect \\
         \hline 
         
         \multirow{2}{*}{RAG\_25\%}&Corpus containing 75\% MPI source code with at least \\
          & one known defect and 25\% known defect-free MPI source code\\ 
         \hline 

        \multirow{2}{*}{RAG\_0\%}&Corpus containing only MPI source code with \\ 
        &each document containing at least one known defect\\ 
    
    \end{tabular}
    \end{adjustbox}
    \medskip
    \label{tab:rag_nomenclatures}
\end{table}


We extended our Few-Shot prompts by incorporating retrieval from our domain-specific databases, adding explicit retrieval roles and contextual information, to systematically measure how varying distributions of correct and defective MPI code influence ChatGPT’s defect detection accuracy across all experimental configurations \cite{Salemi_2024_RAG_Evaluation}.

\begin{framed}
\textbf{\underline{Retrieval Augmented Generation Prompt:}} 

$role\colon user, content\colon \textit{ [ENGINEERED PROMPT] }$

$role\colon assistant, content\colon \textit{[RETRIEVAL RESULTS]}$

\end{framed}

For each of the 241 pre-processed programs in our testing dataset, we performed a retrieval operation and integrated the resulting data into the \textit{[RETRIEVAL RESULTS]} section of the assistant prompt. The Few-Shot prompts were replaced with \textit{[ENGINEERED PROMPT]}. This process was repeated for all 241 test programs, resulting in the construction of 241 unique retrieval-augmented Few-Shot prompts.

\begin{figure}[htbp]
    \centering
    \includegraphics[width=0.6\textwidth]{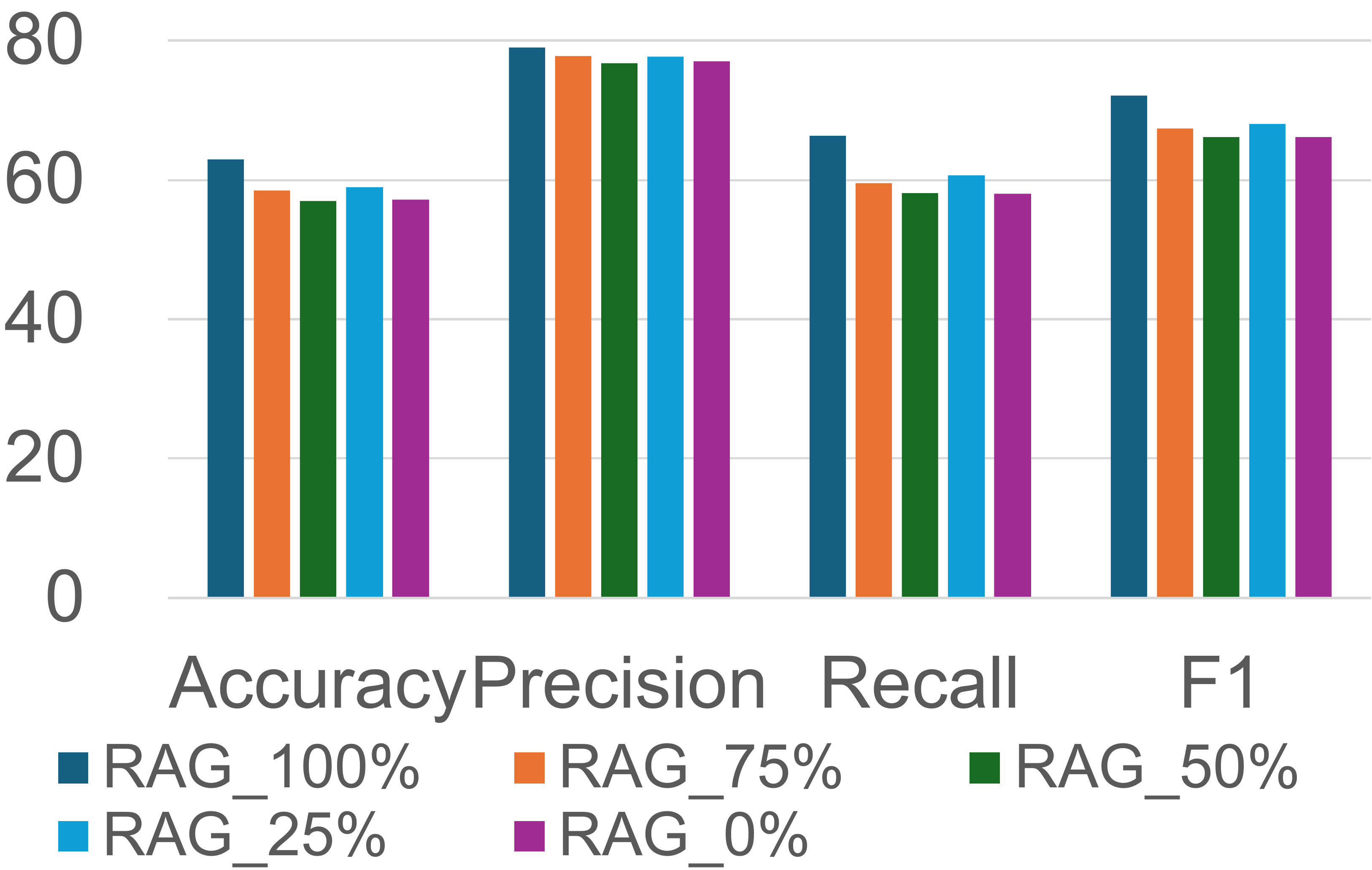}
    \caption{Comparison among different RAG. As the blue bar is the highest, RAG\_100\% is the best.}
    \label{fig:gptrag}

\end{figure}


Our results show that although RAG improves defect detection accuracy relative to the Zero-Shot baseline, it does not match the performance of the Few-Shot+CoT bug-referencing approach (Table~\ref{tab:ablation_details_summary}). In Chroma, retrieval scores are reported as embedding-space distances, where smaller values indicate higher semantic similarity; across all five retrieval databases, scores ranged narrowly from 0.119176 to 0.129195 (Table~\ref{tab:RAG_RETRIEVAL_SCORES}). This tightly bounded, low-distance range indicates that retrieved candidates are highly relevant, stable, and not driven by noise or embedding drift, confirming consistently high-quality retrieval across all retrieval database compositions.


We attribute the limited effectiveness of RAG to several factors: defective code samples often lack explicit annotations, which can bias the model toward labeling plausible-looking but faulty code as correct \cite{RAG_FAILURE_POINTS_barnett2024sevenfailurepointsengineering, RAG_ATTACK_10.1145/3769082}. Moreover, because source code semantics are not well captured by token-level similarity, nearest-neighbor retrieval can produce misleading analogies, and RAG’s sensitivity to chunking and retrieval noise further degrades performance. This issue is particularly acute in MPI, where superficially similar calls may differ in communicators, ordering, or tags \cite{RAG_ATTACK_10.1145/3769082}. These effects are illustrated in Figure~\ref{tab:RAG_RETRIEVAL_SCORES}, with a complete summary of ChatGPT trials provided in Table~\ref{tab:appendix_chatgpttrials}.

\begin{table}[ht]
    \centering
    \caption{ChromaDB Average Retrieval Scores by Retrieval Database. The observed range is narrow and lies in the low-distance regime, indicating all retrieved candidates are tightly clustered around the query vector and therefore semantically highly relevant.} 
    \begin{adjustbox}{max width=0.48\linewidth}
    \begin{tabular}{l|c}  
        \textbf{Retrieval Database}&\textbf{Retrieval Score}\\
        \Xhline{1pt}
        RAG\_100\% & 0.119315 \\
         \hline 
        RAG\_75\% & 0.119244 \\
         \hline 
        RAG\_50\% & 0.119176 \\
         \hline 
        RAG\_25\%  &  0.123624 \\
         \hline 
        RAG\_0\% &   0.129195 \\
    \end{tabular}
    \end{adjustbox}
    \medskip
    \label{tab:RAG_RETRIEVAL_SCORES}
\end{table}

\subsection{Statistical Validation}

Our bug referencing technique exhibits a substantial and statistically significant shift in defect detection behavior relative to the zero shot baseline. Across all five trials, true positives increased from 303 to 851, while false negatives decreased from 567 to 19, yielding an increase in recall from 0.35 to 0.98 and indicating near-complete detection of positive instances. Our improvement is accompanied by a reduction in true negatives from 232 to 87 and a corresponding increase in false positives from 103 to 248, resulting in a decrease in specificity from 0.69 to 0.26. Despite this trade-off, precision remained stable (0.75 to 0.77), and overall accuracy increased from 0.44 to 0.78, with the F1 score improving from 0.47 to 0.86. McNemar’s test confirms that the redistribution of errors toward false positives in exchange for a substantial reduction in false negatives is highly significant $\mathbf{(x^2\approx 233.2, p  < 0.001)}$ . These results demonstrate our bug referencing technique prioritizes defect detection while maintaining precision.

Despite our substantial gains in recall, our bug referencing technique exhibits a marked reduction in specificity, which decreased from 0.69 to 0.26. This increase in false positives implies that a larger number of correct MPI programs are flagged as potentially defective, which may introduce additional analysis or review overhead. In practice, however, false positives in MPI defect detection are often less costly than false negatives, as they can typically be filtered through lightweight static analysis, developer inspection, or secondary validation tools.

\subsection{Evaluation of non-LLM tools}
We compared our approach against established non-LLM, state-of-the-art MPI defect detection tools—CIVL \cite{CIVL_LUO_2017}, ISP \cite{Vakkalanka_2008_ISP_ModelChecking}, MPI-SV \cite{Chen_2020_MPI_SV}, and PARCOACH \cite{PARCOACH2020} which employ static analysis, dynamic analysis, symbolic execution, and model checking respectively (Table~\ref{tab:nonllmtools}). While these tools target key MPI error classes, each exhibits notable limitations, including lack of support for non-blocking operations (CIVL), state-space explosion (ISP), limited deadlock detection (MPI-SV), and high false-positive rates (PARCOACH) \cite{Chen_2020_MPI_SV, MPI_BI, PARCOACH2020}. Because these methods are deterministic, each was evaluated using a single trial on our dataset. As shown in Tables~\ref{tab:nonllmtools} and \ref{tab:rationsandrates}, all four tools produced higher false-positive and false-negative rates than our bug-referencing approach, which achieved a 10\% lower false-positive rate than the best non-LLM tool and a miss-rate of only 2.3\%, approximately 15× lower than the non-LLM average (34.75\%). These results demonstrate that our approach substantially reduces spurious warnings and missed defects, improving MPI defect coverage, developer efficiency, and trust.

\begin{table}[htbp]
    \centering
    \footnotesize
    \setlength{\tabcolsep}{4pt}
    \caption{False Positive and False Negative Ratios. The best results are in \textbf{bold}.}
    \begin{tabular}{l|cc|cc} 
          \multirow{2}{*}{\textbf{Tool}}&\multirow{2}{*}{\textbf{FP}}&\multirow{2}{*}{\textbf{FN}}&\textbf{FP Ratio}& \textbf{FN Ratio}\\ 
              &  &  & \textbf{(Fall-out \%)} & \textbf{(Miss-rate \%)} \\  \Xhline{1pt} 
           CIVL\cite{CIVL_LUO_2017}&61&101& 60.40& 62.35\\ \hline
           ISP\cite{Vakkalanka_2008_ISP_ModelChecking}&94&14& 70.15& 10.77\\ \hline 
           MPISV\cite{Chen_2020_MPI_SV}&64&94& 61.54& 58.02\\ \hline 
           PARCOACH\cite{PARCOACH2020}&79&12& 66.39& 7.89\\ \hline 
           \textbf{Few-Shot+CoT (ours)}&\textbf{50}&\textbf{4}& \textbf{55.56}& \textbf{2.30}\\  
    \end{tabular}
    \label{tab:rationsandrates}
\end{table}

\section{Automated MPI Program Repair}

We evaluated ChatGPT’s ability to automatically repair MPI programs using our defect-detection enhancements, specifically Few-Shot bug references with Chain-of-Thought (CoT) reasoning, and assess whether each generated repair resolves the original defect. Our dataset, drawn from \cite{MPI_BI}, targets MPI error classes that state-of-the-art detectors frequently miss, including local concurrency errors common in one-sided non-blocking communication \cite{MPI_BI}, parameter-matching defects that can induce deadlock, request lifecycle ordering violations, and resource leaks due to unreleased resources. To construct the repair benchmark, we selected defective programs from the original dataset that ChatGPT consistently flagged as defective across all five Few-Shot+CoT trials (true positives), resulting in 164 programs. We then performed a single repair trial by inserting each program into our repair prompt (replacing \textit{[CODE]}) and evaluating the correctness of the proposed repairs.

\begin{framed}
\textbf{\underline{Repair Prompt:}} 
\textit{[Few-Shot Bug Reference Prompt]} + \textit{[Chain of Thought]} + \textit{[CODE]}
You are a Message Passing Interface (MPI) expert in parallel programming. The MPI program above contains errors. Repair the MPI C program and include only the repaired program in your answer. 
\end{framed}



\subsection{MPI Setup}
Testing of ChatGPT's MPI program repairs was performed on an MPICH-based cluster comprised of 4 compute nodes each with 28 CPU cores and 512 GB of RAM running Red-Hat Enterprise Linux.

\subsection{Repair Evaluation Metrics}

We evaluated ChatGPT’s repairs of 164 MPI programs using three criteria: successful compilation, removal of resource leaks, and absence of deadlocks; a repair was considered successful only if all three conditions were satisfied, with any failure classified as an unsuccessful repair. Because manual inspection was impractical given the dataset size and execution complexity, we developed an automated evaluation suite to test all repaired programs generated using our repair prompt. Deadlocks were detected by injecting a 900-second \texttt{alarm} call into each program, with executions exceeding this limit recorded as deadlocked, while resource leaks were identified by running each program under \textit{valgrind} with \texttt{--leak-check=yes} via \textit{mpirun} and analyzing the captured outputs.

\subsection{Repair Analysis}
Because our work is the first to measure ChatGPT's ability to detect and repair defects in MPI programs, there is no established baseline metric for repair of the programs in our dataset. 



\textbf{Successful compilation.} We evaluated ChatGPT’s repair capability by measuring whether the generated programs compiled successfully, finding that 12 of 164 repair attempts failed to compile, corresponding to a 92.68\% successful compilation rate. Analysis of the compiler outputs revealed that five failures were due to newly introduced multiple variable declarations within the same scope, four resulted from syntax errors caused by missing statement terminators, and three arose from incorrect argument types passed to MPI communication routines. Because these defects were not present in the original programs, we attribute the latter class of errors to hallucinations and unintended code modifications introduced by the model during the repair process.



\textbf{Resource Leak Removal}. We evaluated ChatGPT’s ability to repair resource leaks on the subset of 13 programs with known leaks from our dataset of 164 repaired programs, initially observing a high failure rate. Further analysis showed that many reported leaks originated in MPI libraries rather than in the application code, prompting us to exclude MPI library leaks to align with our goal of assessing repairs to MPI programs themselves. After this adjustment, ChatGPT successfully repaired 11 of the 13 programs, corresponding to an 84.76\% success rate, with one failure attributable to a syntactic error in the generated C code (a missing semicolon).

\textbf{Deadlock Removal}. Because deadlocks are a common error in distributed parallel programming using MPI \cite{HUang_2020_Predictive_Deadlock}, our experimental results for deadlock removal are of particular interest and value to the HPC community. Overall, ChatGPT successfully repaired 76.83\% of the deadlock defects in our dataset. 

\begin{figure}
    \centering
    \includegraphics[width=0.7\textwidth]{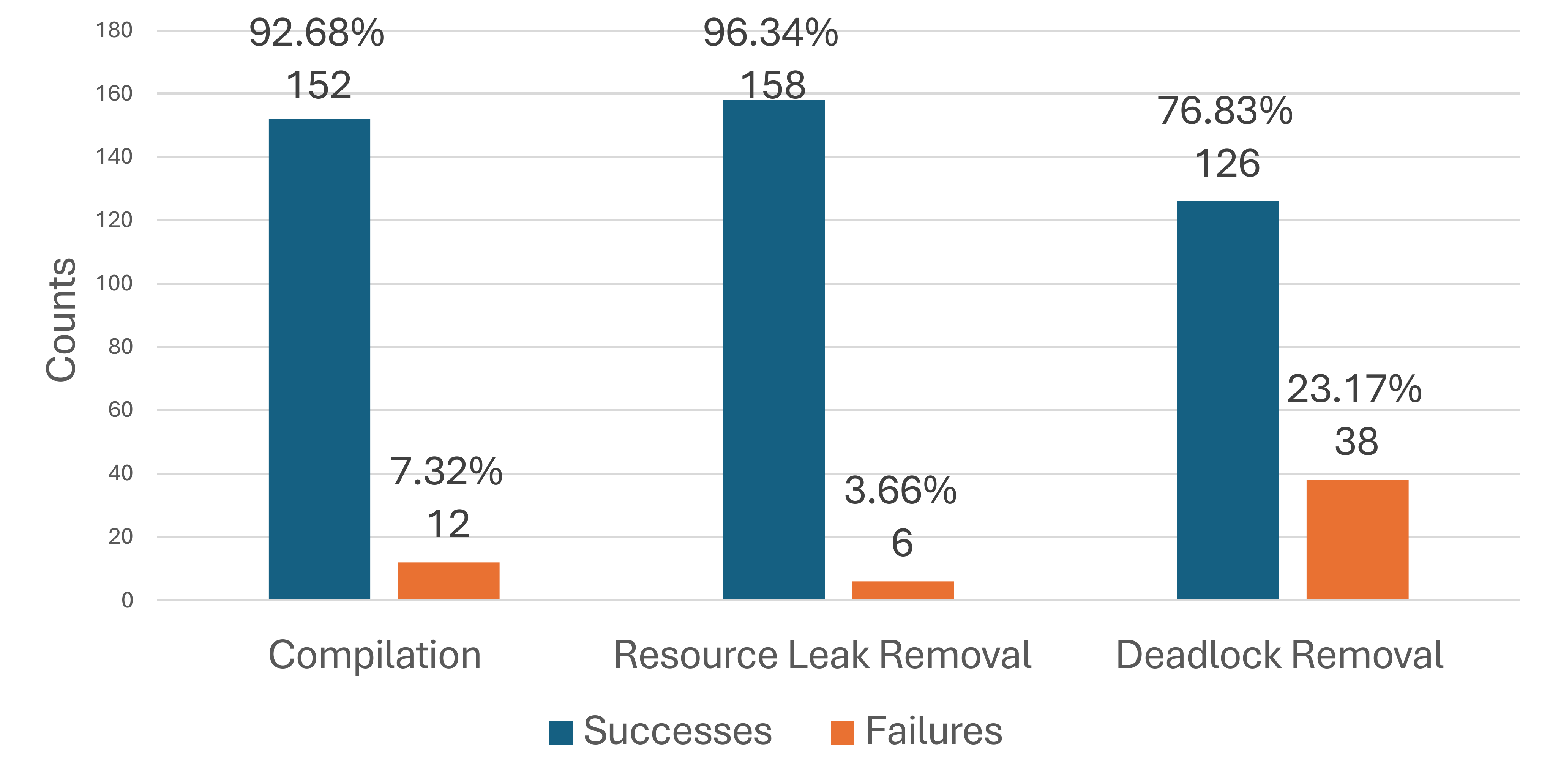}
    \caption{Distribution of Repair Successes and Failures by Evaluation Metric including a summary of ChatGPT repair success rates across different evaluation criteria: Successful Compilation, Resource Leak Removal, and Deadlock Removal. Depicting the specific areas of strength and opportunities for future enhancement in MPI program repair.}
    \label{fig:RepairSuccessesVsFailurebyMetric}
\end{figure}

\textbf{Repair Success}. Using our metric for classifying a repair as successful, our experimental results indicate that 125 repairs of our dataset containing 164 programs met all three of our successful inclusion criteria. Our measured success rate for utilizing ChatGPT to successfully repair MPI programs in our dataset is 76.21\% (125 of 164 repair successes) .   Figure \ref{fig:RepairSuccessesVsFailurebyMetric} illustrates our success rates vs failures for each type of defect that we attempted to repair using ChatGPT.

\textbf{Repair Failures.}
ChatGPT’s primary repair limitation was its inability to resolve deadlock defects, a common and challenging class of MPI errors. To identify contributing factors, we analyzed repair successes and failures by defect type (Figure \ref{fig:RepairSuccessesVsFailurebyMetric}) and found that most failures stem from parameter-matching defects, which typically manifest as execution deadlocks (IE: deadlock removal failures). Listing~\ref{lst:MistatchedTagDefect} illustrates the difficulty of repairing this class of MPI error.  The specific challenge in repairing the defect shown in Listing \ref{lst:MistatchedTagDefect} lies in the need for a level of understanding beyond basic syntax checking rules. Detecting this error requires comprehension of the intent behind the MPI library calls and recognition that both types and values of communication parameter tags must align correctly.

\section{Generality and Applicability to Other Large Language Models}
Last, we investigate the external validity of our bug-referencing approach on large language models other than ChatGPT. We selected three LLMs for our external validity study: Llama2, Code Llama, and Qwen2.5-coder.  LLama2 is an open-source large language model selected to contrast the results obtained with ChatGPT. Code Llama and Qwen2.5-coder are code-specific models chosen to evaluate our approach using models developed specifically for coding tasks  \cite{codellama_rozière2024codellamaopenfoundation} \cite{QWEN2.5_coder_hui2024qwen25codertechnicalreport} 

We selected Llama2 due to its demonstrated competitiveness with proprietary models (such as ChatGPT), its responsible and open release model, its no-cost licensing model,  and its availability in a 13 billion parameter size which matches our ChatGPT GPT-3.5 Turbo model's parameter size \cite{llama2_touvron2023llama2openfoundation}. Code Llama was selected due to its state-of-the-art performance among open models for programming tasks, large context window, zero-shot performance, and its availability in a 13 billion parameter size which matches our ChatGPT model size \cite{codellama_rozière2024codellamaopenfoundation}.  Qwen2.5-coder was selected due to its permissive licensing, adoption by developers in real-world applications, proficiency in handling a variety of programming languages, performance in code reasoning and repair, and its availability in a 14 billion parameter size which closely matches our ChatGPT model size of 13 billion parameters \cite{QWEN2.5_coder_hui2024qwen25codertechnicalreport}.

\subsection{Experimental Setup}
We performed the same experimental trials for Llama2, Code Llama, and Qwen2.5-coder which we performed in our ChatGPT GPT-3.5 Turbo trials. We deployed all three models on a Ubuntu 22.04 Linux System with two Nvidia 4070 Super GPUs running Ollama. All three LLMs were used in their default configuration (we made no modifications to parameter sizes or context window lengths). We recorded all experimental trials, by model and prompting technique for Llama2, Code Llama, and Qwen2.5-coder and recorded them in \ref{tab:appendix_b_llama2trials}, \ref{tab:appendix_c_codellamatrials}, and \ref{tab:appendix_d_qwen2.5trials} respectively. 

\subsection{Classification Performance Metrics}

\begin{figure}[htbp]
\centering
\makebox[\textwidth][c]{%
    \begin{subfigure}{0.36\textwidth}
        \includegraphics[width=\linewidth]{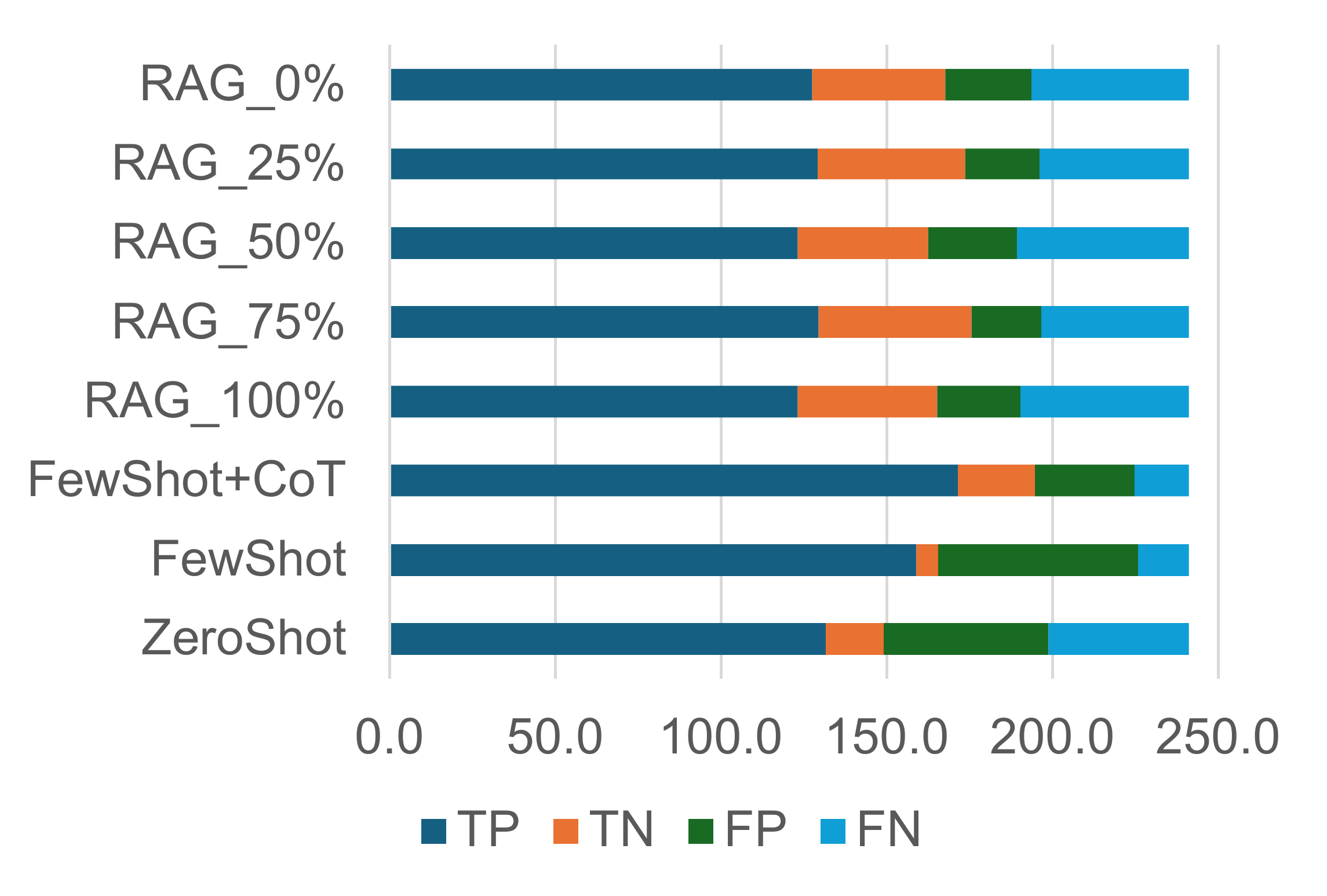}
        \caption{Llama2}
        \label{fig:llama2_bar}
    \end{subfigure}
    \hspace{-1.2em} 
    \begin{subfigure}{0.36\textwidth}
        \includegraphics[width=\linewidth]{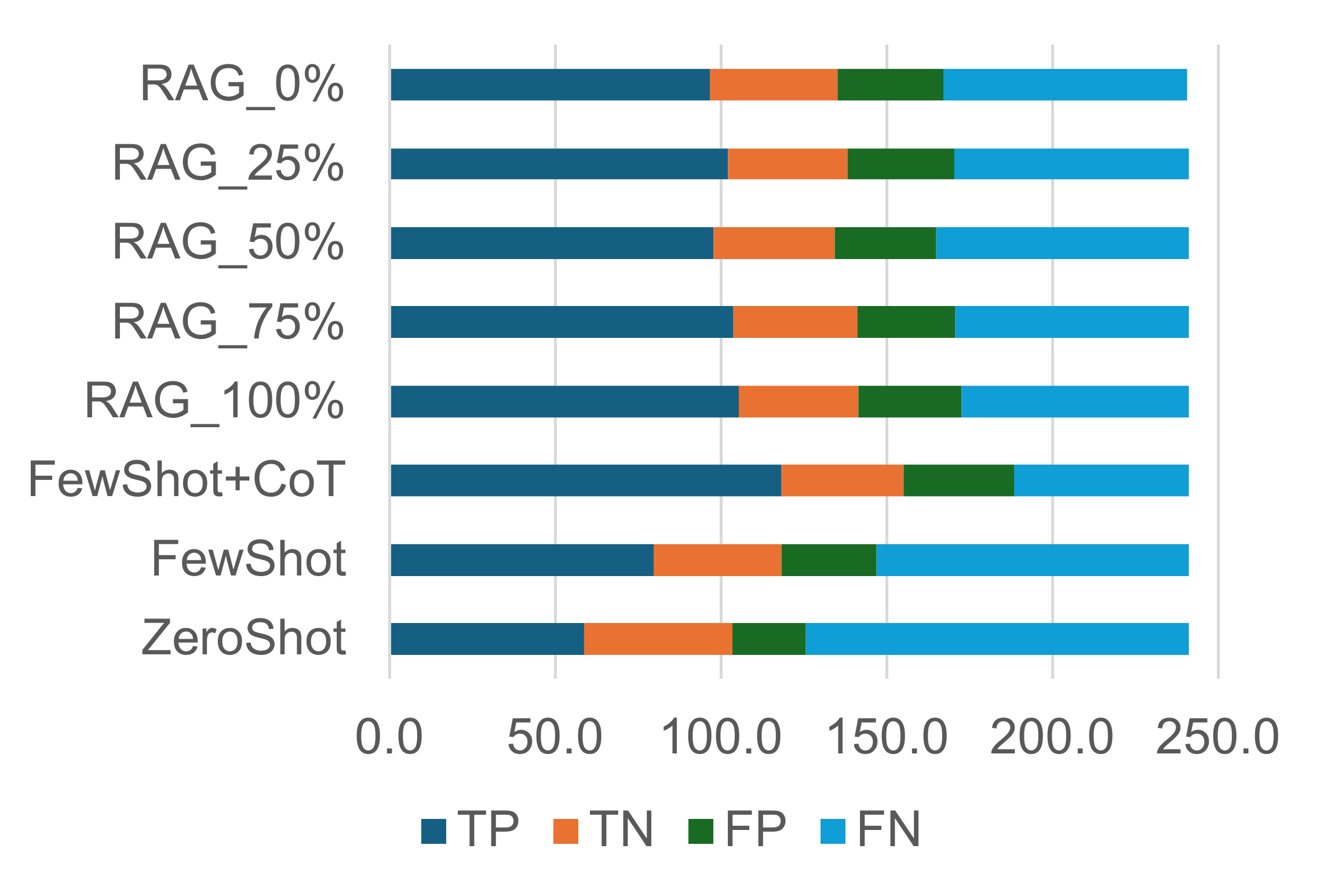}
        \caption{Code Llama}
        \label{fig:codellama_bar}
    \end{subfigure}
    \hspace{-1.2em}
    \begin{subfigure}{0.36\textwidth}
        \includegraphics[width=\linewidth]{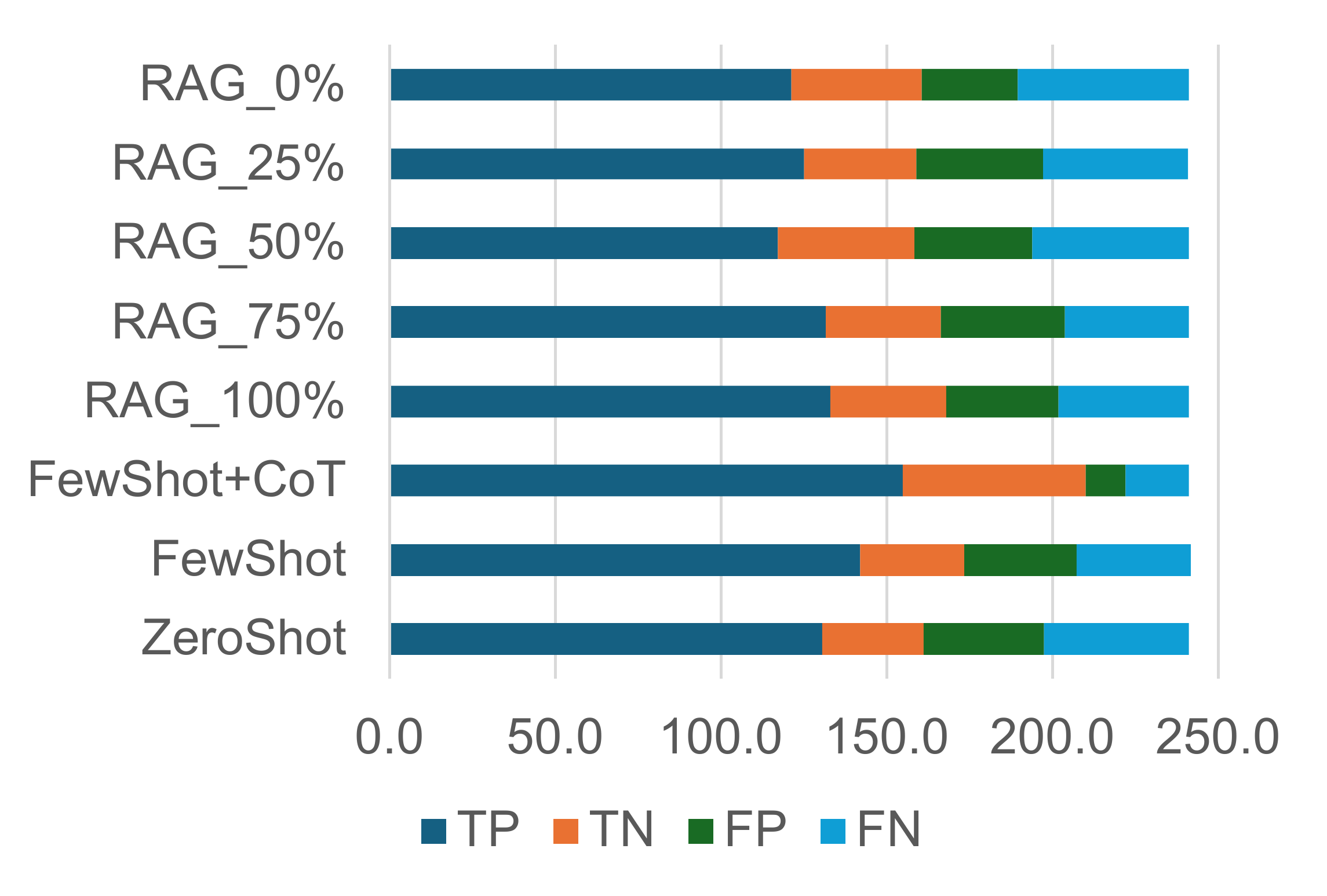}
        \caption{Qwen2.5-coder}
        \label{fig:qwen_bar}
    \end{subfigure}
}
\caption{Comparison of Zero-Shot, Few-Shot, Few-Shot+CoT and Few-Shot+CoT+RAG True Positive(TP), True Negative(TN), False Positive(FP), and False Negative(FP) results. The inclusion of our bug referencing Few-Shot+CoT reasoning exhibits the largest improvement in true positives and reduction in false negatives across all three LLMs.
}
\end{figure}

Our bug referencing technique, when applied to all three models, demonstrates the same improvements over each model's baseline MPI defect detection performance. As depicted in Figures \ref{fig:llama2_bar}, \ref{fig:codellama_bar}, and \ref{fig:qwen_bar} our bug referencing approach provides not only the largest improvement in true positives, but also the largest reduction in false negatives when compared to the other  prompting techniques in our study, clearly proving the generality of our bug referencing approach.

\begin{figure}[htbp]
    \centering
    \begin{subfigure}{0.32\textwidth}
        \centering
        \includegraphics[width=\textwidth]{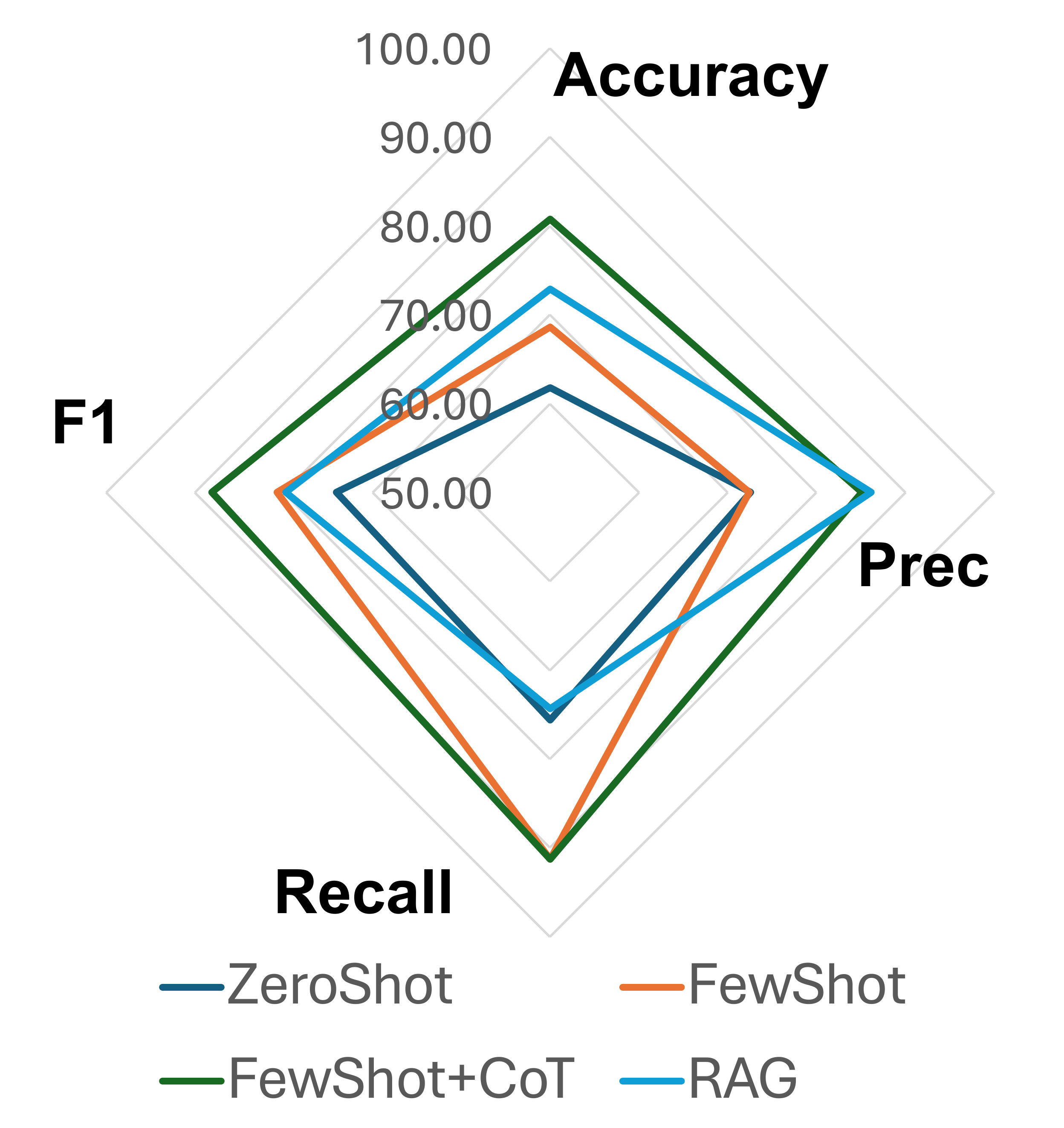}
        \caption{Llama2}
        \label{fig:radar_zero_vs_fs_fscot_llama2}
    \end{subfigure}
    \hfill
    \begin{subfigure}{0.32\textwidth}
        \centering
        \includegraphics[width=\textwidth]{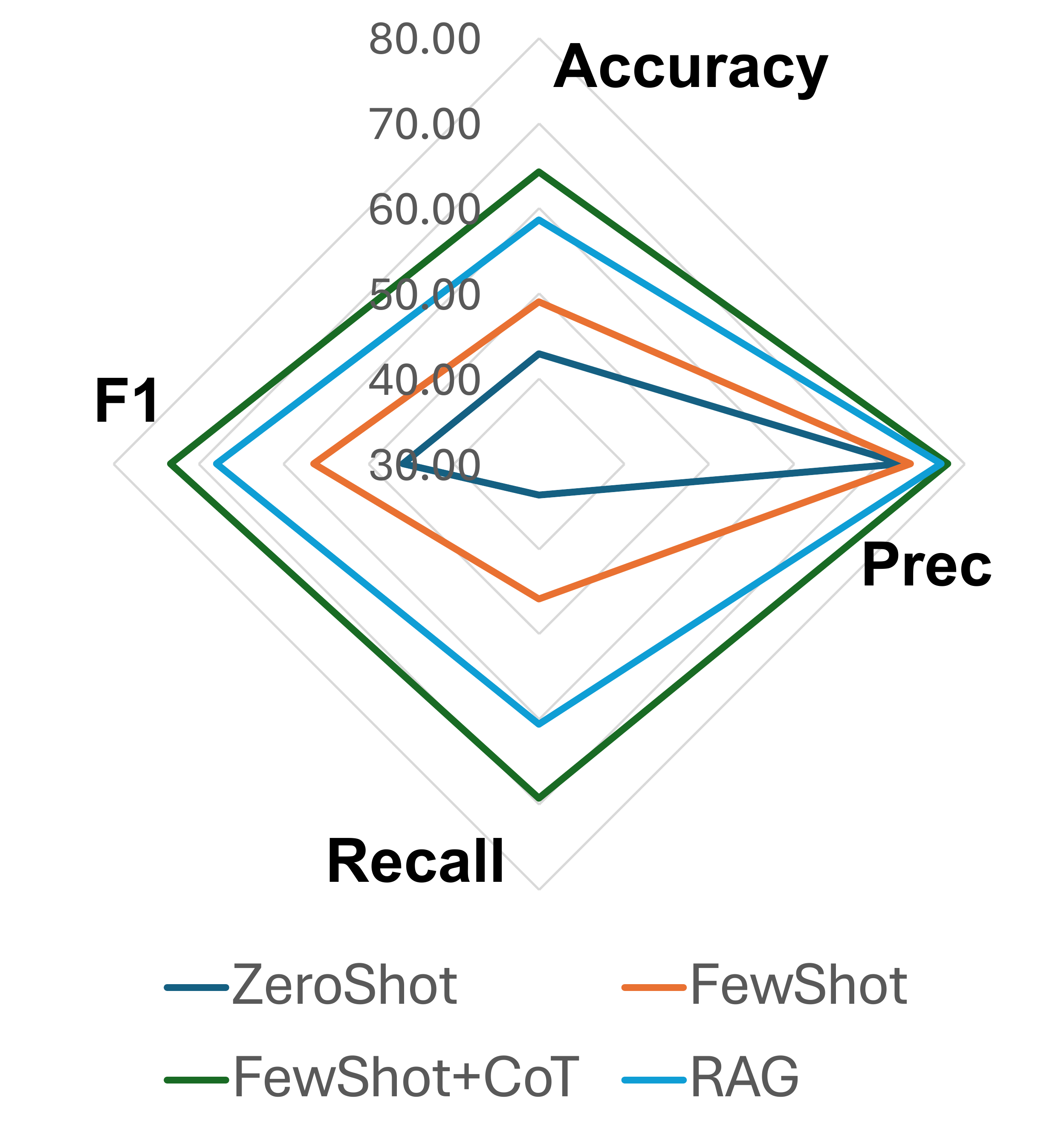}
        \caption{Code Llama}
        \label{fig:radar_zero_vs_fs_fscot_codellama}
    \end{subfigure}
    \begin{subfigure}{0.32\textwidth}
        \centering
        \includegraphics[width=\textwidth]{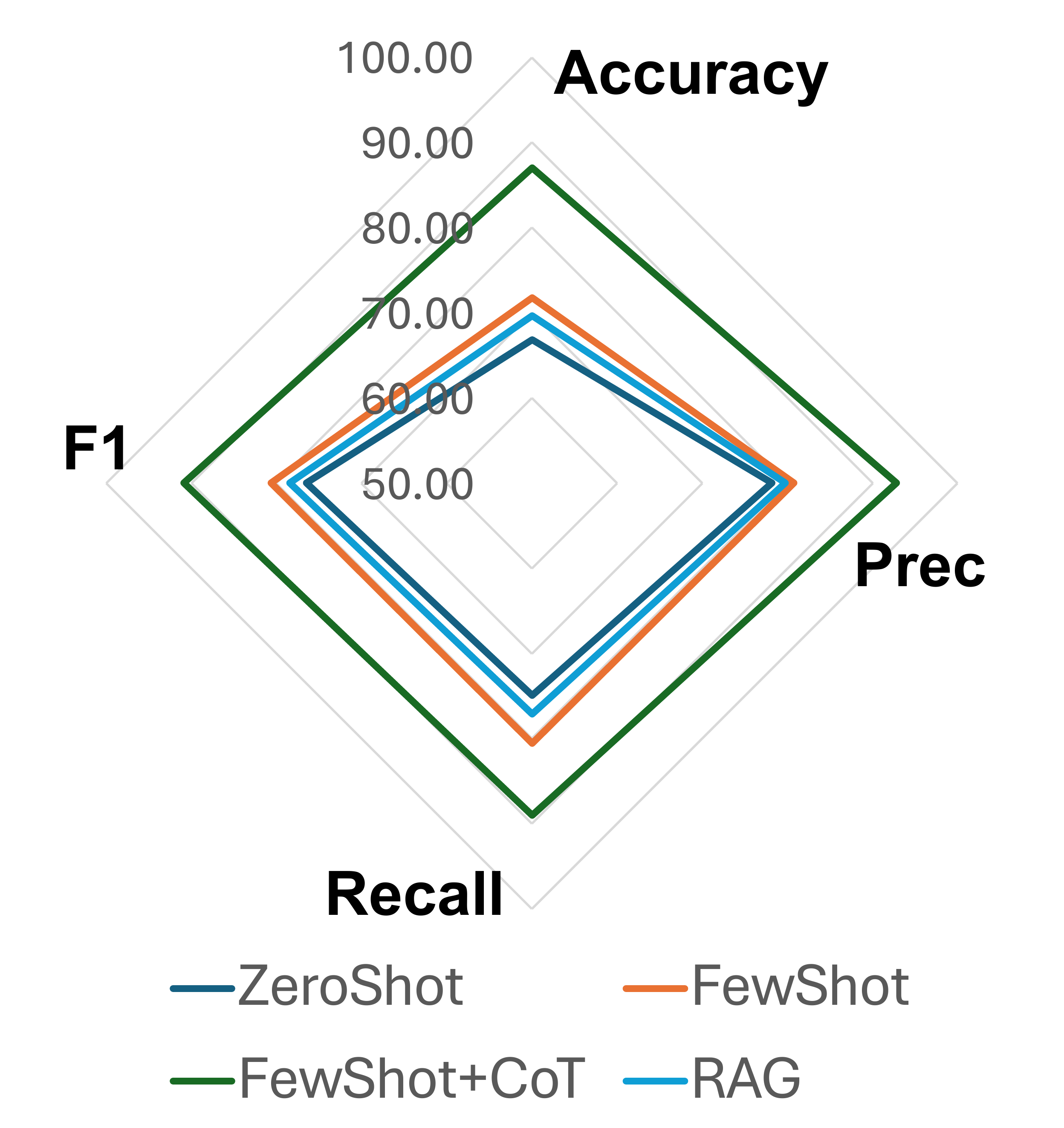}
        \caption{Qwen2.5-coder}
        \label{fig:radar_zero_vs_fs_fscot_qwen}
    \end{subfigure}
    \caption{Comparison of Zero-Shot, Few-Shot, Few-Shot+CoT and Few-Shot+CoT+RAG prompting strategies based on accuracy, precision, recall, and F1-score. Our bug referencing technique provides the best classification and detection metrics across all three LLMs }
    \label{fig:radar_zero_vs_fs_fscot}
\end{figure}

Our bug referencing approach provides the largest improvement in true positives across all models while simultaneously delivering the largest decrease in false negatives. Congruently, our bug referencing approach delivery the best improvements in accuracy, precision, recall and F1 MPI defect detection scores across all three models, as depicted in Tables \ref{fig:radar_zero_vs_fs_fscot_llama2},  \ref{fig:radar_zero_vs_fs_fscot_codellama}, and \ref{fig:radar_zero_vs_fs_fscot_qwen}. These improvements in evaluation metrics further demonstrate the generality of our bug referencing approach.

\subsection{Statistical Validity}

Our bug referencing technique demonstrates a statistically significant improvement over the zero shot baseline when used with all three LLMs. We performed McNemar’s test on our results from all three LLMs comparing the zero-shot baseline to our bug referencing technique for each model. We recorded the results for McNemar's test in Table \ref{tab:variances_by_model} for each of the three models. The results in Table \ref{tab:variances_by_model} demonstrate that our bug referencing technique achieves genuine improvement in discriminative capability rather than a simple shift in decision threshold across all three models. 

\begin{table}[ht]
    \centering
    \caption{McNemar's Test Results when comparing zero shot baseline to our bug referencing approach across all LLMs. The results demonstrate the statistical significance of our bug referencing approach}
    \begin{adjustbox}{max width=0.4\linewidth}
    \begin{tabular}{l|c|c}     \textbf{Model}&\textbf{$\mathbf{x^2}$} &\textbf{p} \\ 
        \Xhline{1pt}                   
         Llama2  &  ${\approx 64.1} $ & ${< 0.001}$ \\
         \hline
         Code Llama &${\approx 164.1}$ & ${< 0.001}$ \\
         \hline
         Qwen2.5-Coder&${\approx 181.2}$ & ${< 0.001}$ \\
    \end{tabular}
    \end{adjustbox}
    \medskip  \label{tab:variances_by_model}
\end{table}

\subsection{Cross-LLM Performance Stability}

We calculated and recorded variance ($\sigma^2$) of True Positives, True Negatives, False Positives and False Negatives by prompting technique across all three LLMs. A comparison of these variances demonstrates similar trends noted in our ChatpGPT trials: our bug referencing approach has the lowest variance in True Positives, True Negatives, False Positives, and False Negatives across all models and prompting techniques. These lower variances in our bug referencing technique continue the trend observed in our ChatGPT trials, further demonstrating the higher reliability, stability, , less non-determinism, and improved model reasoning of our approach.

\begin{table}[htbp]
  \centering
  \caption{Comparison of Variances $\sigma^2$ among Llama2, Code Llama, Qwen2.5-coder prompting techniques. Again, our integrated bug referencing approach produces the lowest variance for True Positives (TP), True Negatives (TN), False Positives (FP), and False Negatives (FN). CoT=Chain Of Thought Reasoning.}
  \begin{adjustbox}{max width=\linewidth}
    \begin{tabular}{l|rrrr|rrrr|rrrr}
    \multicolumn{1}{c|}{\multirow{2}[1]{*}{\textbf{Prompting Tech}}} & \multicolumn{4}{c|}{\textbf{Llama2 $\sigma^2$}} & \multicolumn{4}{c|}{\textbf{Code Llama $\sigma^2$}} & \multicolumn{4}{c}{\textbf{Qwen2.5-coder $\sigma^2$}} \\
          & {\textbf{TP}} & {\textbf{TN}} & {\textbf{FP}} & {\textbf{FN}} & {\textbf{TP}} & {\textbf{TN}} & {\textbf{FP}} & \multicolumn{1}{l|}{\textbf{FN}} & {\textbf{TP}} & {\textbf{TN}} & {\textbf{FP}} & {\textbf{FN}} \\
    \midrule
    Zero-Shot  & 33.3  & 4.3   & 4.3   & 33.3  & 29.3  & 31.7  & 31.7  & 29.3  & 34.8  & 5.3   & 5.3   & 34.8 \\
    Few-Shot  & 14.7  & 2.3   & 2.3   & 14.7  & 25.7  & 4.3   & 4.3   & 25.7  & 46.7  & 11.3  & 29.7  & 36.8 \\
    \textbf{Few-Shot+CoT}  & \textbf{3.3}   & \textbf{0.7}   & \textbf{1.0}   & \textbf{3.3}   & \textbf{3.7}   & \textbf{3.2}   & \textbf{3.8}   & \textbf{10.3}  & \textbf{1.7}   & \textbf{1.5}   & \textbf{0.5}   & \textbf{0.7} \\
    Few-Shot+CoT+RAG\_100\%  & 9.7   & 2.5   & 2.5   & 9.7   & 48.8  & 5.5   & 5.5   & 48.8  & 10.5  & 5.2   & 11.7  & 4.3 \\
    Few-Shot+CoT+RAG\_75\%  & 65.3  & 6.7   & 6.7   & 65.3  & 41.3  & 4.3   & 4.3   & 41.3  & 47.3  & 10.3  & 11.3  & 43.3 \\
    Few-Shot+CoT+RAG\_50\%   & 40.7  & 27.2  & 21.7  & 38.7  & 29.8  & 9.3   & 7.3   & 32.7  & 8.5   & 4.8   & 19.8  & 15.7 \\
    Few-Shot+CoT+RAG\_25\%   & 36.0  & 7.3   & 7.3   & 36.0  & 46.0  & 18.7  & 66.2  & 37.3  & 39.5  & 15.7  & 51.7  & 100.2 \\
    Few-Shot+CoT+RAG\_0\%  & 38.3  & 7.2   & 35.5  & 23.8  & 74.8  & 21.3  & 43.0  & 73.3  & 16.7  & 9.2   & 12.5  & 52.8 \\
    \end{tabular}%
  \end{adjustbox}
  \label{tab:variances}%
\end{table}%

\subsection{Defect Repair Performance}

\begin{figure}[htbp]
\centering
\makebox[\textwidth][c]{%
    \begin{subfigure}{0.36\textwidth}
        \includegraphics[width=\linewidth]{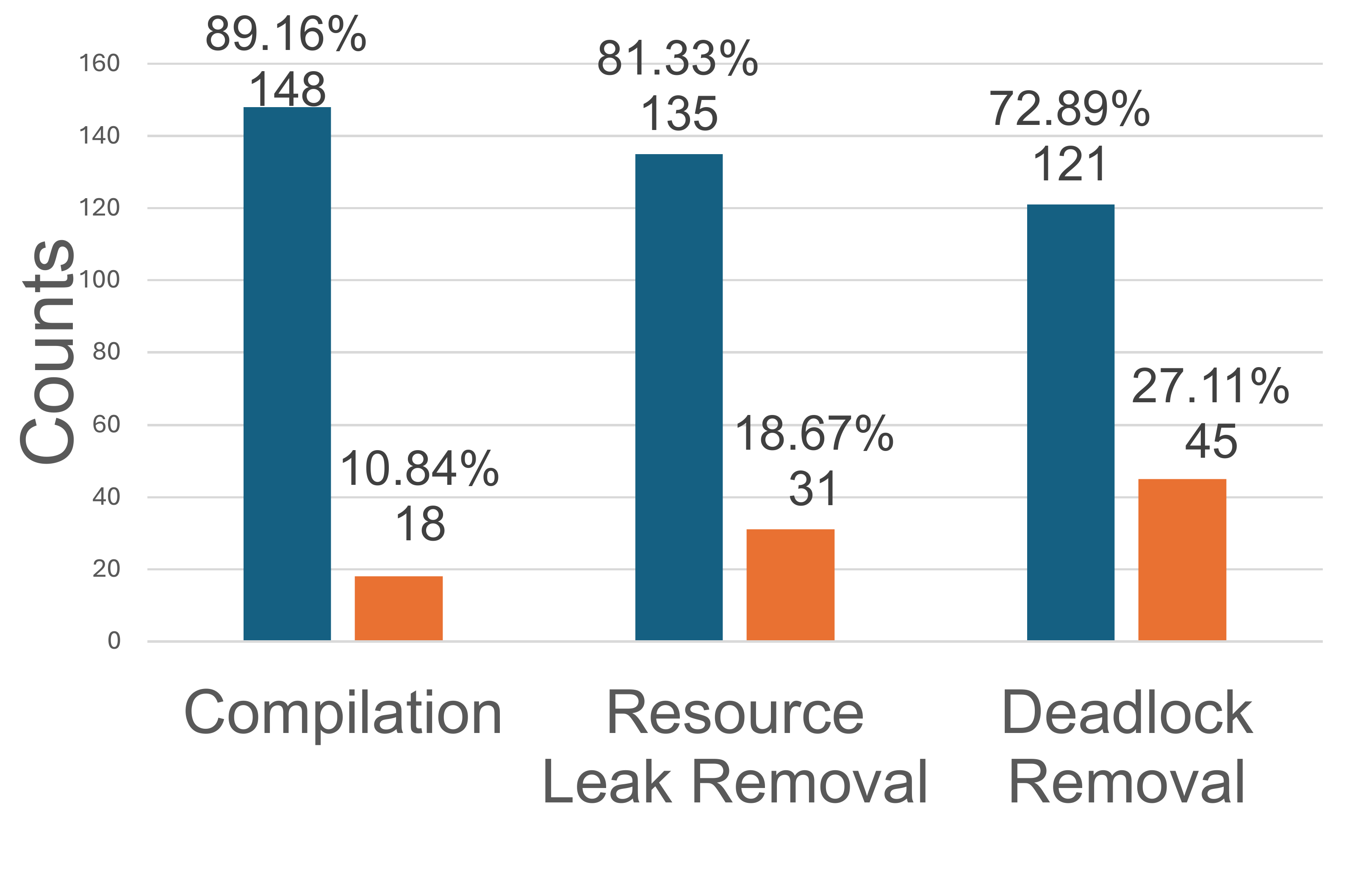}
        \caption{Llama2}
        \label{fig:llama2_RepairSuccessesVsFailurebyMetric}
    \end{subfigure}
    \hspace{-1.2em} 
    \begin{subfigure}{0.36\textwidth}
        \includegraphics[width=\linewidth]{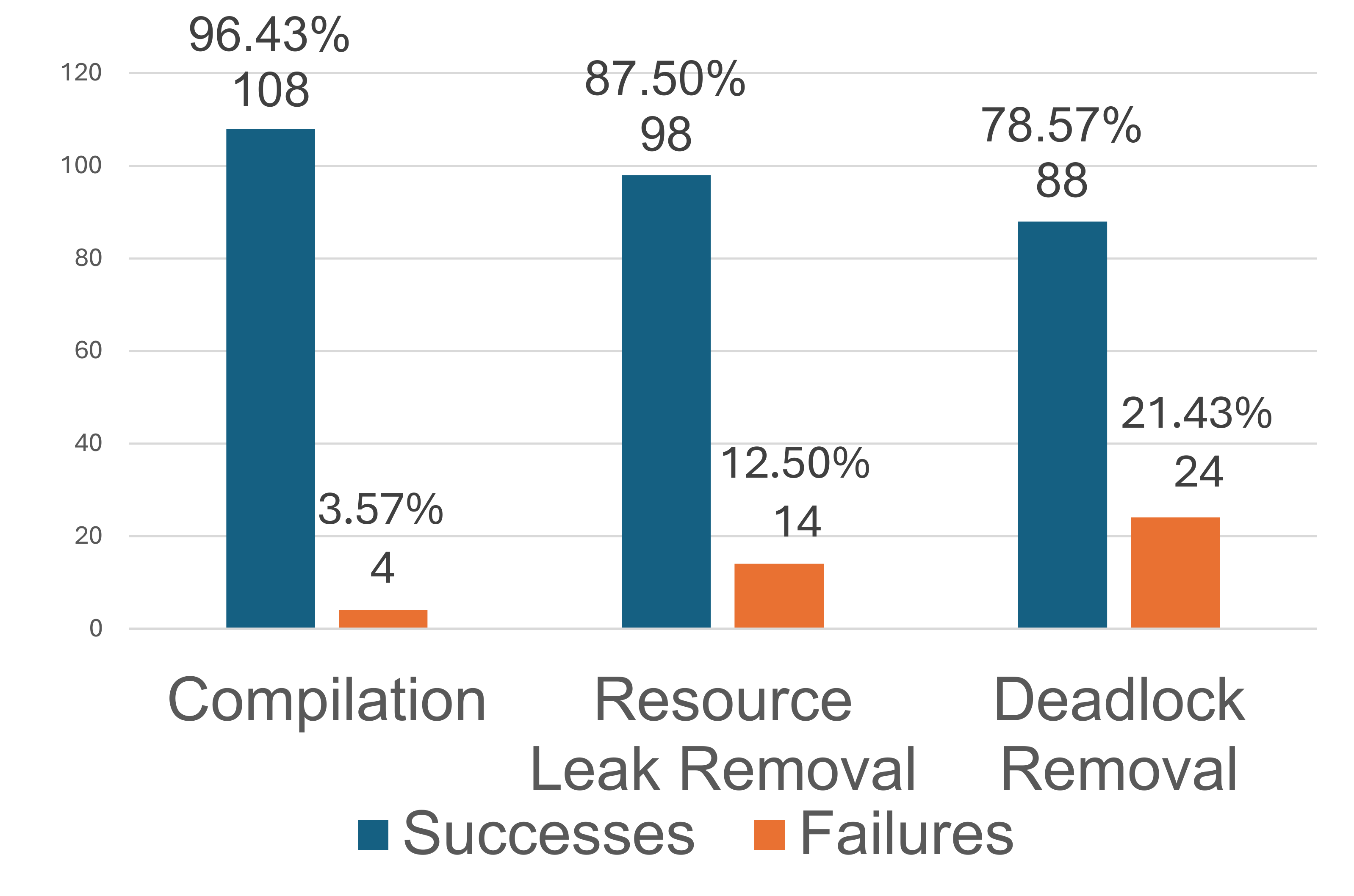}
        \caption{Code Llama}
        \label{fig:codellama_RepairSuccessesVsFailurebyMetric}
    \end{subfigure}
    \hspace{-1.2em}
    \begin{subfigure}{0.36\textwidth}
        \includegraphics[width=\linewidth]{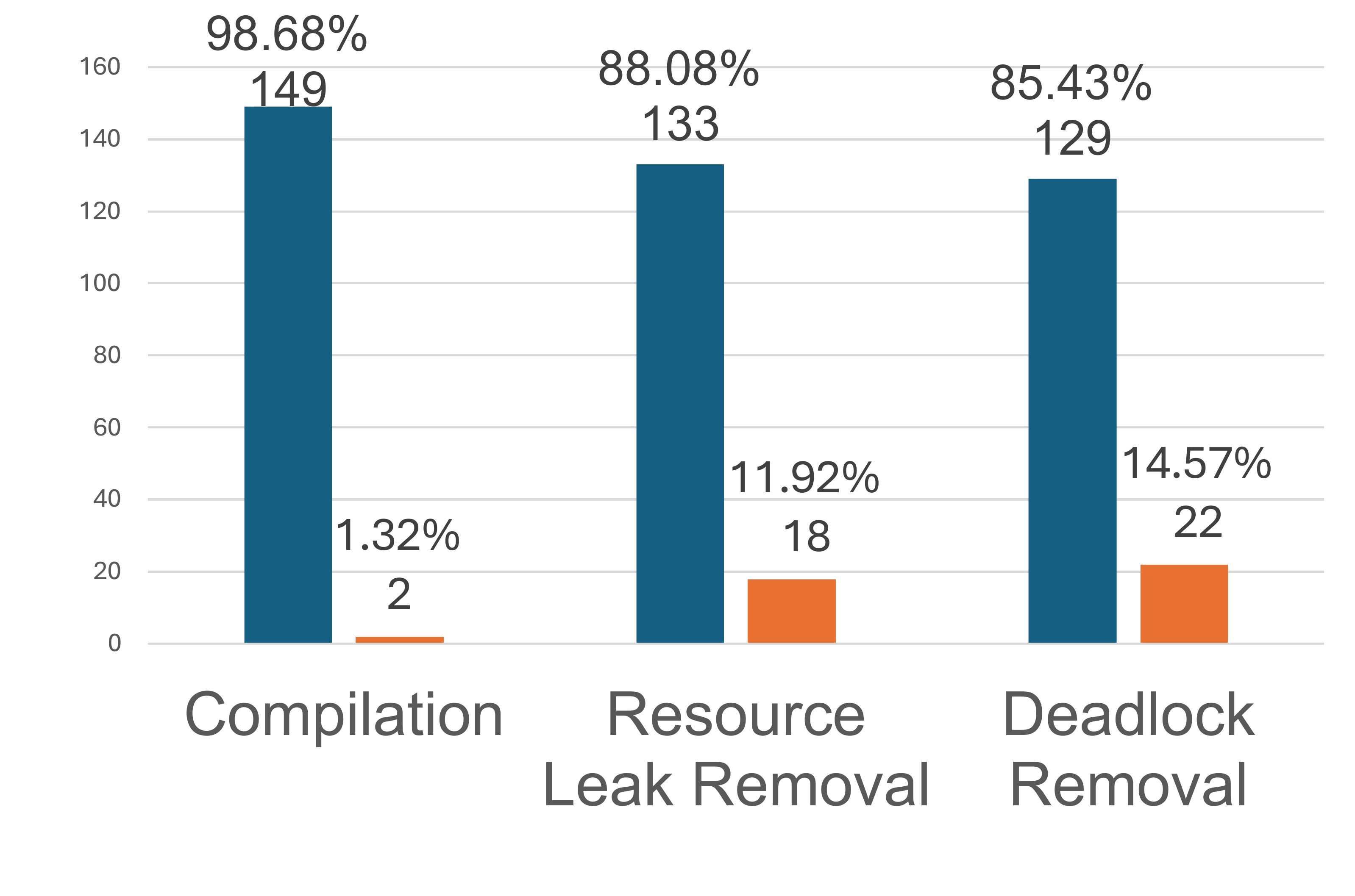}
        \caption{Qwen2.5-coder}
        \label{fig:qwen_RepairSuccessesVsFailurebyMetric}
    \end{subfigure}
}
\caption{Repair success and failure rates by defect type across all three LLMs.}
\end{figure}

We repeated the repair program construction and selection methodology used in our ChatGPT repair study for each of the three LLMs.  The repair success rates by defect type for Llama2, Code Llama, and Qwen2.5-coder using our bug referencing approach are depicted in Figures  \ref{fig:llama2_RepairSuccessesVsFailurebyMetric}, \ref{fig:codellama_RepairSuccessesVsFailurebyMetric}, and  \ref{fig:qwen_RepairSuccessesVsFailurebyMetric}. We noted similar trends in successes vs. failures by metric for all three LLMs that were noted in our initial repair trial using ChatGPT. It can be noted from Figure \ref{fig:qwen_RepairSuccessesVsFailurebyMetric} that Qwen2.5-coder produces repairs with the highest successful compilation rates, and resource leak and deadlock removal.  We attribute these results to Qwen2.5-Coder's strong capability in detecting programming errors and suggesting plausible repairs due to its targeted fine-tuning on large-scale, code-centric corpora that emphasize syntactic correctness, API usage patterns, and common bug–fix transformations \cite{QWEN2.5_coder_hui2024qwen25codertechnicalreport}. Furthermore, Qwen2.5-coder's training dataset composition, which includes diverse programming languages and real-world code–repair examples, enables the model to generate contextually appropriate, implementation-level fixes with high consistency \cite{QWEN2.5_coder_hui2024qwen25codertechnicalreport}.

\subsection{Summary}

 The results from our external validity study demonstrate our bug referencing technique is robust and applicable to other LLMs in several study areas. First, our bug referencing technique outperformed the zero shot baseline and few shot performance for all three LLMs, duplicating the results of our bug referencing technique from our initial ChatGPT trials. Second, our bug referencing approach provides the best accuracy, precision, and recall in MPI defect detection on our dataset when used with Llama2, Code Llama, and Qwen2.5-coder LLMs as depicted in Figures \ref{fig:radar_zero_vs_fs_fscot_llama2}, \ref{fig:radar_zero_vs_fs_fscot_codellama}, and \ref{fig:radar_zero_vs_fs_fscot_qwen} respectively. Furthermore, McNemar's test performed on baseline vs bug referencing technique results across all LLMs demonstrate the statistical significant of our approach and that our improvements are not due to random fluctuations. Last, our bug referencing approach provides the lowest variances in true positives, true negatives, false positives, and false negatives demonstrating the superior model reasoning provided by our technique. 

 The addition of retrieval augmented generation using corpora comprised of varying distributions of good vs. defective MPI code degraded the performance of our approach across Llama2, Code Llama, and Qwen2.5-coder. This effect duplicates the effects noted in our ChatGPT trials and is depicted in Figures \ref{fig:radar_zero_vs_fs_fscot_llama2}, \ref{fig:radar_zero_vs_fs_fscot_codellama}, and \ref{fig:radar_zero_vs_fs_fscot_qwen}. We attribute these repeated trends from our ChatGPT retrieval experiments to the constraints noted in our previous analysis. Defective program samples (like those in 80\% of our retrieval corpora) typically lack explicit defect annotations or explanatory source code comments, causing retrieved code to appear plausible and biasing all three models toward incorrect defect classifications due to missing contextual signals \cite{RAG_FAILURE_POINTS_barnett2024sevenfailurepointsengineering,RAG_ATTACK_10.1145/3769082}. Additionally, because source code is not natural language, token-level similarity does not imply semantic equivalence, making nearest neighbor retrieval and chunking brittle and prone to misleading analogies \cite{RAG_ATTACK_10.1145/3769082}. This issue is especially pronounced in MPI programs where identical call names may conceal critical differences in communicators, ordering, and/or tags.

\begin{figure}[htbp]
    \centering
    \begin{subfigure}{0.32\textwidth}
        \centering
        \includegraphics[width=\textwidth]{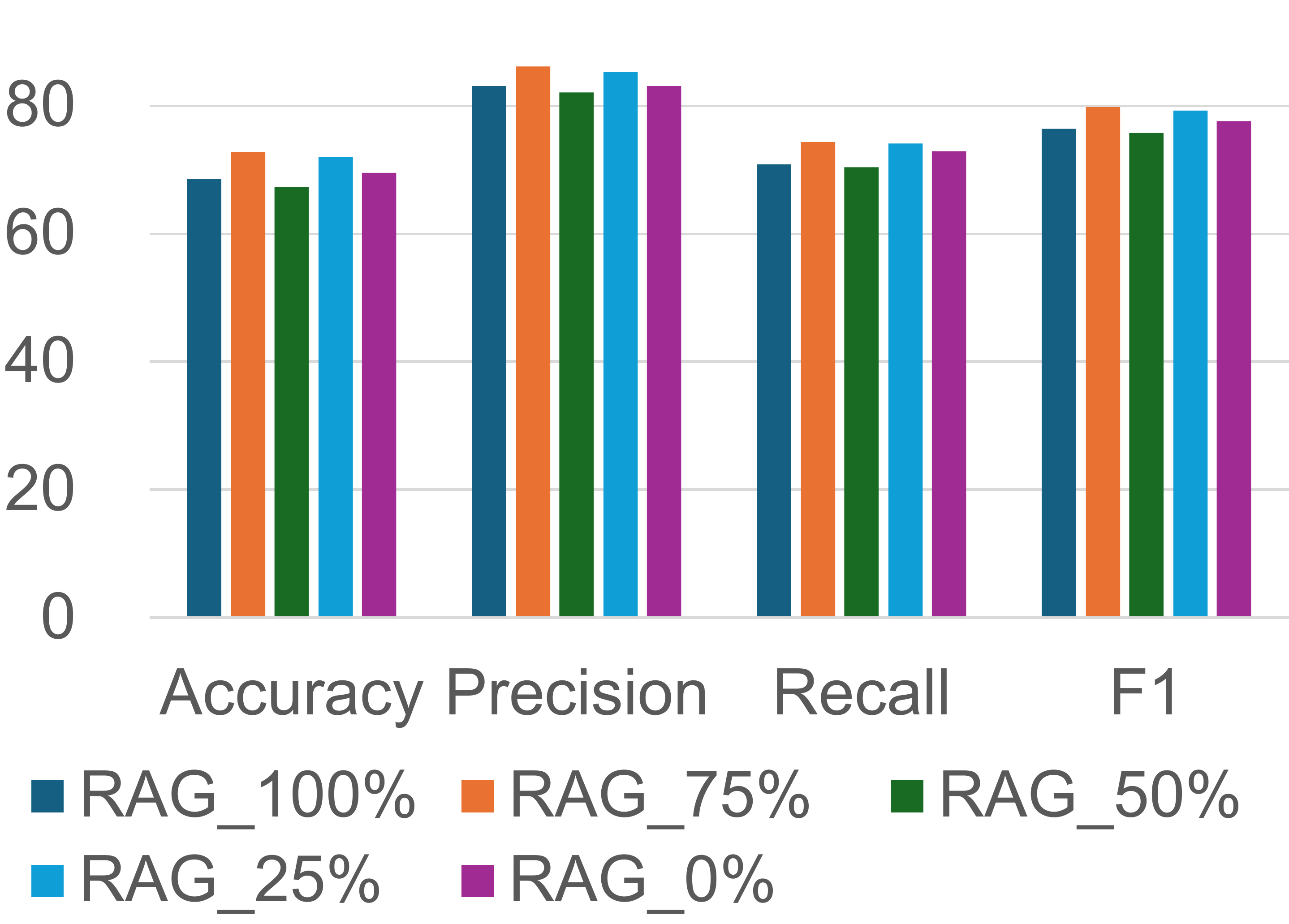}
        \caption{Llama2}
        \label{fig:rag_by_corpus_llama2}
    \end{subfigure}
    \hfill
    \begin{subfigure}{0.32\textwidth}
        \centering
        \includegraphics[width=\textwidth]{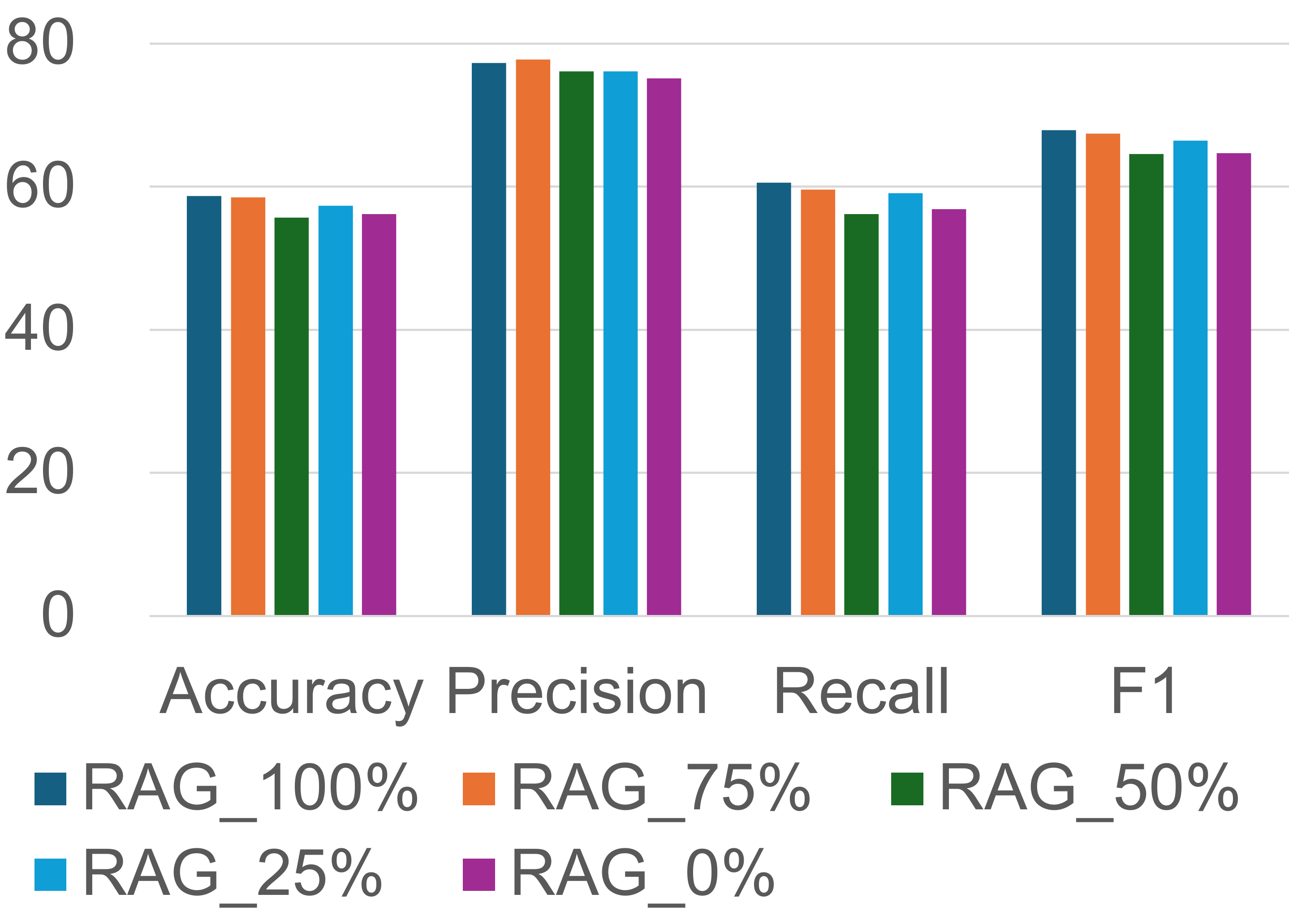}
        \caption{Code Llama}
        \label{fig:rag_by_corpus_codellama}
    \end{subfigure}
    \begin{subfigure}{0.32\textwidth}
        \centering
        \includegraphics[width=\textwidth]{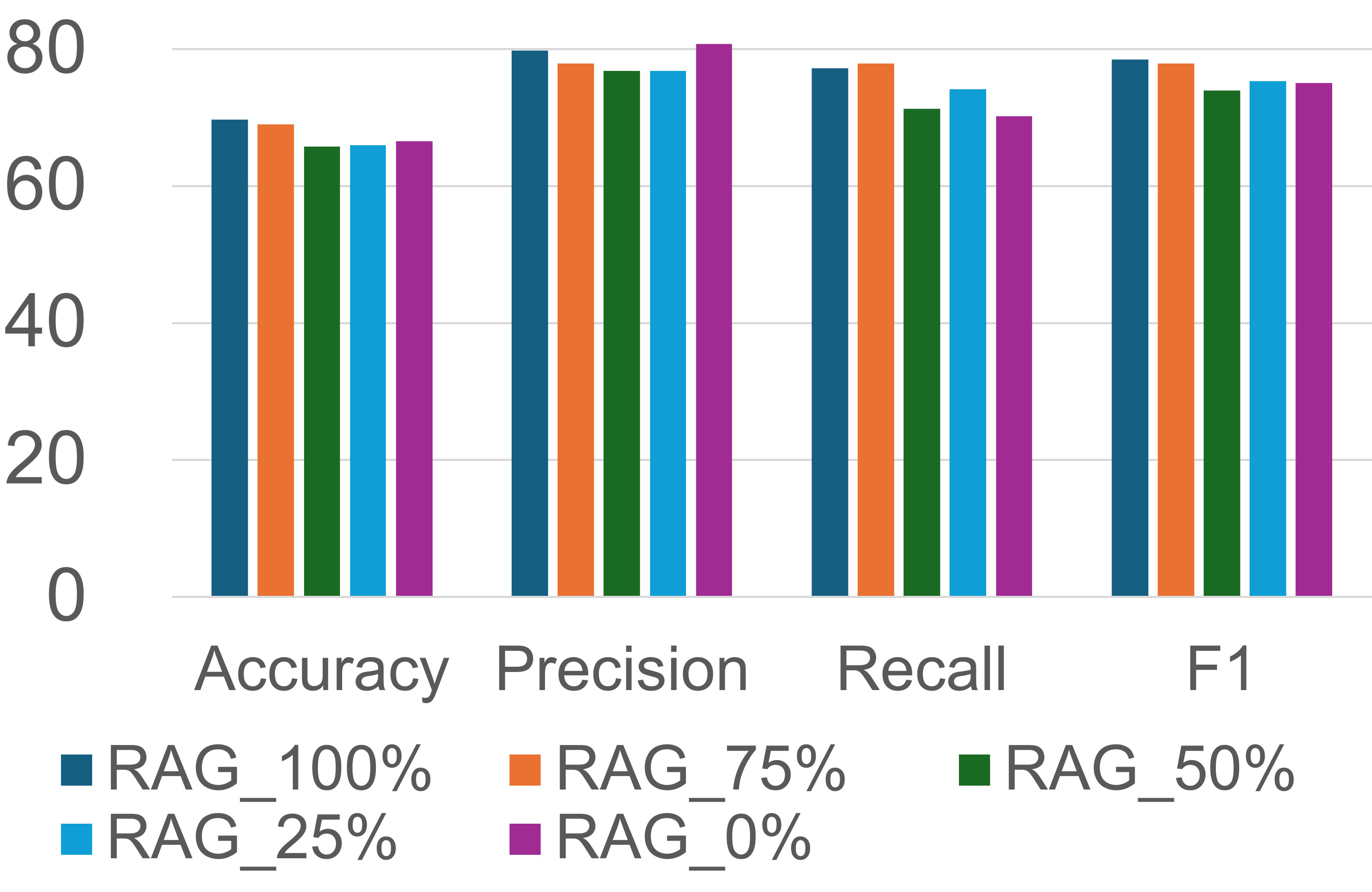}
        \caption{Qwen2.5-coder}
        \label{fig:rag_by_corpus_qwen}
    \end{subfigure}
    \caption{Accuracy, Precision, Recall, and F1 Score by Retrieval Corpus Composition using Llama2, Code Llama, and Qwen2.5-coder, showing repeated trends for each retrieval corpus by LLM.}
    \label{fig:rag_by_corpus_all}
\end{figure}
 
Figure \ref{fig:rag_by_corpus_all} illustrates similar performance trends across LLama2 (Figure \ref{fig:rag_by_corpus_llama2}), Code Llama (Figure \ref{fig:rag_by_corpus_codellama}), and Qwen-2.5coder (Figure \ref{fig:rag_by_corpus_qwen}), indicating consistent retrieval across all three LLMs. Again, the observed range of 0.119176 to 0.129195 in our ChromaDB retrieval distances is narrow and in the low-distance regime, indicating retrieved candidates are narrowly clustered around the query vector for all LLMs in our study. The small spread $\mathbf{(\approx 0.01)}$ further implies stable retrievals not dominated by noise or embedding drift.

\section{Threats to Validity}
\subsection{Internal}
The primary internal threat to our validity stems from the inherent characteristics and internal operation of all LLMs: non-determinism. The primary issue with non-determinism in LLMs is its tendency to produce varying outputs for the same input. We mitigated this risk by performing multiple experimental trials using the same prompts and input data for each trial and reporting the average of each trial. Additionally, because ChatGPT was trained on publicly available source code, we cannot guarantee that the ChatGPT model training data did not include programs from our MPI test dataset. This issue can be addressed if the exact composition of ChatGPT's training dataset is known; however, this information is not made available by OpenAI. We mitigated this risk by selecting programs from varying sources when constructing our MPI dataset that was used for all of the LLMs in our study. 

\subsection{External}
We evaluated ChatGPT on a publicly available dataset comprised of known good and defective MPI programs. Our findings may not generalize to other datasets, programming paradigms, or LLMs. Additionally, the ChatGPT GPT-3.5 Turbo model used may not remain publicly available as OpenAI may choose to discontinue its availability at a future date, posing a risk to long-term repeatability.

The open-source, permissive-licensed LLMs that were used undergo continuous retraining and improvements. The exact models used in our trial will experience training updates, refinements, and eventual replacement posing risks to both short-term and long-term repeatability.

\section{Conclusions}

Our bug-referencing approach substantially improves ChatGPT’s MPI defect detection accuracy over its out-of-the-box configuration, with domain-expert–guided Chain-of-Thought reasoning yielding further gains and the largest reduction in variance across true and false positives and negatives, demonstrating strong generalization to other LLMs. Given the difficulty of MPI error detection and the scalability and usability limitations of existing static and dynamic tools, our results show that LLMs can effectively identify nuanced MPI defects that traditional approaches often miss. In repair tasks, ChatGPT successfully corrected 84.62\% of resource leaks and 83.44\% of deadlocks, with lower performance on mismatched parameter tag defects (67.39\%), and these results generalized across multiple open-source and code-focused models. Overall, by compensating for the lack of defective MPI examples in LLM training data through Few-Shot bug references and Chain-of-Thought reasoning, our approach significantly improves detection of hard-to-find MPI errors in datasets where state-of-the-art tools exhibit high false-negative rates.

\section{Future Work}

Although our results demonstrate substantial improvements in MPI defect detection using Few-Shot Learning with Chain-of-Thought reasoning and domain guidance, LLMs frequently failed to correctly repair deadlock defects, motivating future work on prompts that more effectively convey MPI program intent during repair. Additionally, our RAG experiments show that retrieval database composition has minimal impact on detection accuracy, highlighting limitations of text-based embeddings in capturing MPI concurrency, message-passing semantics, and communication hazards. These findings suggest the need for code-aware representations and retrieval methods tailored to MPI-specific program structure, building on the effectiveness of domain-expert guidance observed in this study.


\bibliographystyle{elsarticle-num} 
\bibliography{ref}

\appendix
\newpage 
\section{Experimental Trial Results for ChatGPT}
\label{sec:appendix_a}
\renewcommand{\thetable}{\thesection.\arabic{table}} 
\setcounter{table}{0} 

{\footnotesize
\setlength{\tabcolsep}{0.2pt}

\sisetup{
  detect-weight=true,
  detect-inline-weight=math,
  table-number-alignment=center
}

\begin{xltabular}{\textwidth}{l|S S S S S |S S S S}
\caption{Experimental Trial Results Using ChatGPT}
\label{tab:appendix_chatgpttrials}\\

\toprule
\textbf{Study} & \textbf{Trial} & \textbf{TP} & \textbf{TN} & \textbf{FP} & \textbf{FN}
& \textbf{Acc(\%)} & \textbf{Prec(\%)} & \textbf{Rec(\%)} & \textbf{F1(\%)} \\
\midrule
\endfirsthead

\toprule
\textbf{Study} & \textbf{Trial} & \textbf{TP} & \textbf{TN} & \textbf{FP} & \textbf{FN}
& \textbf{Acc(\%)} & \textbf{Prec(\%)} & \textbf{Rec(\%)} & \textbf{F1(\%)} \\
\midrule
\endhead

\midrule
\multicolumn{10}{r}{\textit{Continued on next page}}\\
\endfoot

\bottomrule
\endlastfoot
        Zero-Shot & 1 & 56    & 46    & 21    & 118   & 42.32 & 72.72 & 32.18 & 44.62   \\
        Zero-Shot & 2 & 55    & 42    & 25    & 119   & 40.24 & 68.75 & 31.60 & 43.30   \\
        Zero-Shot & 3 & 65    & 45    & 22    & 109   & 45.64 & 74.71 & 37.35 & 49.80   \\
        Zero-Shot & 4 & 63    & 48    & 19    & 111   & 46.05 & 76.82 & 36.20 & 49.21   \\
        Zero-Shot & 5 & 64    & 51    & 16    & 110   & 47.71 & 80.00   & 36.78 & 50.39   \\
        \textbf{Zero-Shot} & \textbf{Avg} & \textbf{60.6} & \textbf{46.4} & \textbf{20.6} & \textbf{113.4} & \textbf{44.39} & \textbf{74.60} & \textbf{34.82} & \textbf{47.46}  \\
        \midrule 
        Few-Shot & 1 & 121   & 12    & 59    & 49    & 55.18 & 67.22 & 71.17 & 69.14   \\
        Few-Shot & 2 & 124   & 11    & 51    & 55    & 56.01 & 70.85 & 69.27 & 70.05   \\
        Few-Shot & 3 & 119   & 15    & 58    & 49    & 55.60 & 67.23 & 70.83 & 68.98   \\
        Few-Shot & 4 & 121   & 17    & 55    & 48    & 57.26 & 68.75 & 71.59 & 70.14   \\
        Few-Shot & 5 & 124   & 15    & 53    & 49    & 57.67 & 70.05 & 71.67 & 70.85   \\
        \textbf{Few-Shot} & \textbf{Avg} & \textbf{121.8} & \textbf{14.0} & \textbf{55.2} & \textbf{50.0} & \textbf{56.34} & \textbf{68.82} & \textbf{70.91} & \textbf{69.83}   \\
        \midrule 
        Few-Shot+CoT & 1 & 172   & 16    & 51    & 2     & 78.00 & 77.13 & 98.85 & 86.64   \\
        Few-Shot+CoT & 2 & 169   & 18    & 49    & 5     & 77.59 & 77.52 & 97.12 & 86.22   \\
        Few-Shot+CoT & 3 & 172   & 16    & 51    & 2     & 78.00 & 77.13 & 98.85 & 86.64   \\
        Few-Shot+CoT & 4 & 167   & 17    & 50    & 7     & 76.34 & 76.95 & 95.97 & 85.42   \\
        Few-Shot+CoT & 5 & 171   & 20    & 47    & 3     & 79.25 & 78.44 & 98.27 & 87.24   \\
        \textbf{Few-Shot+CoT} & \textbf{Avg} & \textbf{170.2} & \textbf{17.4} & \textbf{49.6} & \textbf{3.8} & \textbf{77.84} & \textbf{77.43} & \textbf{97.81} & \textbf{86.43}  \\
        \midrule 
        FS+CoT+RAG\_100\% & 1 & 128   & 35    & 32    & 46    & 67.63 & 80.00   & 73.56 & 76.64  \\
        FS+CoT+RAG\_100\% & 2 & 119   & 34    & 31    & 57    & 63.48 & 79.33 & 67.61 & 73.00  \\
        FS+CoT+RAG\_100\% & 3 & 116   & 34    & 33    & 58    & 62.24 & 77.85 & 66.66 & 71.82  \\
        FS+CoT+RAG\_100\% & 4 & 112   & 39    & 28    & 62    & 62.65 & 80.00   & 64.36 & 71.33  \\
        FS+CoT+RAG\_100\% & 5 & 104   & 38    & 29    & 70    & 58.92 & 78.19 & 59.77 & 67.75  \\
        \textbf{FS+CoT+RAG\_100\%} & \textbf{Avg} & \textbf{115.8} & \textbf{36} & \textbf{30.6} & \textbf{58.6} & \textbf{62.98} & \textbf{79.07} & \textbf{66.39} & \textbf{72.11} \\
        \midrule 
        FS+CoT+RAG\_75\% & 1 & 100   & 37    & 30    & 74    & 56.84 & 76.92 & 57.47 & 65.78  \\
        FS+CoT+RAG\_75\% & 2& 106   & 36    & 31    & 68    & 58.92 & 77.37 & 60.91 & 68.16  \\
        FS+CoT+RAG\_75\% & 3 & 113   & 36    & 31    & 61    & 61.82 & 78.47 & 64.94 & 71.06  \\
        FS+CoT+RAG\_75\% & 4 & 103   & 37    & 30    & 71    & 58.09 & 77.44 & 59.19 & 67.10  \\
        FS+CoT+RAG\_75\% & 5 & 96    & 41    & 26    & 78    & 56.84 & 78.68 & 55.17 & 64.86  \\
        \textbf{FS+CoT+RAG\_75\%} & \textbf{Avg} & \textbf{103.6} & \textbf{37.4} & \textbf{29.6} & \textbf{70.4} & \textbf{58.50} & \textbf{77.77} & \textbf{59.54} & \textbf{67.39} \\
        \midrule 
        FS+CoT+RAG\_50\% & 1 & 96    & 36    & 31    & 78    & 54.77 & 75.59 & 55.17 & 63.78 \\
        FS+CoT+RAG\_50\% & 2 & 99    & 33    & 34    & 75    & 54.77 & 74.43 & 56.89 & 64.49 \\
        FS+CoT+RAG\_50\% & 3 & 101   & 35    & 32    & 73    & 56.43 & 75.93 & 58.04 & 65.79 \\
        FS+CoT+RAG\_50\% & 4 & 106   & 38    & 29    & 68    & 59.75 & 78.51 & 60.91 & 68.60  \\
        FS+CoT+RAG\_50\% & 5 & 104   & 40    & 27    & 70    & 59.75 & 79.38 & 59.77 & 68.19  \\
        \textbf{FS+CoT+RAG\_50\%} & \textbf{Avg} & \textbf{101.2} & \textbf{36.4} & \textbf{30.6} & \textbf{72.8} & \textbf{57.09} & \textbf{76.77} & \textbf{58.16} & \textbf{66.17}  \\
        \midrule 
        FS+CoT+RAG\_25\% & 1  & 115   & 30    & 37    & 59    & 60.16 & 75.65 & 66.09 & 70.55 \\
        FS+CoT+RAG\_25\% & 2 & 107   & 39    & 28    & 67    & 60.58 & 79.25 & 61.49 & 69.25 \\
        FS+CoT+RAG\_25\% & 3 & 104   & 34    & 33    & 70    & 57.26 & 75.91 & 59.77 & 66.88 \\
        FS+CoT+RAG\_25\% & 4 & 109   & 39    & 28    & 65    & 61.41 & 79.56 & 62.64 & 70.09 \\
        FS+CoT+RAG\_25\% & 5 & 93    & 41    & 26    & 81    & 55.60 & 78.15 & 53.44 & 63.48 \\
        \textbf{FS+CoT+RAG\_25\%} & \textbf{Avg} & \textbf{105.6} & \textbf{36.6} & \textbf{30.4} & \textbf{68.4} & \textbf{59.00} & \textbf{77.70} & \textbf{60.68} & \textbf{68.05}  \\
        \midrule 
        FS+CoT+RAG\_0\% & 1 & 97    & 24    & 43    & 77    & 50.20 & 69.28 & 55.74 & 61.78  \\
        FS+CoT+RAG\_0\% & 2 & 110   & 36    & 31    & 64    & 60.58 & 78.01 & 63.21 & 69.84  \\
        FS+CoT+RAG\_0\% & 3 & 93    & 40    & 27    & 81    & 55.18 & 77.50 & 53.44 & 63.26  \\
        FS+CoT+RAG\_0\% & 4 & 94    & 42    & 25    & 80    & 56.43 & 78.99 & 54.02 & 64.16 \\
        FS+CoT+RAG\_0\% & 5 & 111   & 42    & 25    & 63    & 63.48 & 81.61 & 63.79 & 71.61  \\
        \textbf{FS+CoT+RAG\_0\%} & \textbf{Avg} & \textbf{101} & \textbf{36.8} & \textbf{30.2} & \textbf{73} & \textbf{57.17} & \textbf{77.08} & \textbf{58.04} & \textbf{66.13}  \\
   
    \end{xltabular}
}

\section{Experimental Trial Results for Llama2}
\setcounter{table}{0} 
{\footnotesize
\setlength{\tabcolsep}{0.2pt}

\sisetup{
  detect-weight=true,
  detect-inline-weight=math,
  table-number-alignment=center
}

\begin{xltabular}{\textwidth}{l|S S S S S |S S S S}
\caption{Experimental Trial Results Using Llama2}
\label{tab:appendix_b_llama2trials}\\

\toprule
\textbf{Study} & \textbf{Trial} & \textbf{TP} & \textbf{TN} & \textbf{FP} & \textbf{FN}
& \textbf{Acc(\%)} & \textbf{Prec(\%)} & \textbf{Rec(\%)} & \textbf{F1(\%)} \\
\midrule
\endfirsthead

\toprule
\textbf{Study} & \textbf{Trial} & \textbf{TP} & \textbf{TN} & \textbf{FP} & \textbf{FN}
& \textbf{Acc(\%)} & \textbf{Prec(\%)} & \textbf{Rec(\%)} & \textbf{F1(\%)} \\
\midrule
\endhead

\midrule
\multicolumn{10}{r}{\textit{Continued on next page}}\\
\endfoot

\bottomrule
\endlastfoot
        Zero-Shot & 1 & 126   & 19    & 48    & 48    & 60.17 & 72.41 & 72.41 & 72.41 \\
        Zero-Shot & 2 & 138   & 20    & 47    & 36    & 65.56 & 74.59 & 79.31 & 76.88 \\
        Zero-Shot & 3 & 125   & 15    & 52    & 49    & 58.09 & 70.62 & 71.84 & 71.23 \\
        Zero-Shot & 4 & 134   & 16    & 51    & 40    & 62.24 & 72.43 & 77.01 & 74.65 \\
        Zero-Shot & 5 & 135   & 17    & 50    & 39    & 63.07 & 72.97 & 77.59 & 75.21 \\
        \textbf{Zero-Shot} & \textbf{Avg} & \textbf{131.6} & \textbf{17.4} & \textbf{49.6} & \textbf{42.4} & \textbf{61.83} & \textbf{72.61} & \textbf{75.63} & \textbf{74.08} \\
        \midrule 
        Few-Shot & 1 & 154   & 8     & 59    & 20    & 67.22 & 72.30 & 88.51 & 79.59 \\
        Few-Shot & 2 & 164   & 7     & 60    & 10    & 70.95 & 73.21 & 94.25 & 82.41 \\
        Few-Shot & 3 & 157   & 5     & 62    & 17    & 67.22 & 71.69 & 90.23 & 79.90 \\
        Few-Shot & 4 & 158   & 5     & 62    & 16    & 67.63 & 71.82 & 90.80 & 80.20 \\
        Few-Shot & 5 & 161   & 8     & 59    & 13    & 70.12 & 73.18 & 92.53 & 81.73 \\
        \textbf{Few-Shot } & \textbf{Avg} & \textbf{158.8} & \textbf{6.6} & \textbf{60.4} & \textbf{15.2} & \textbf{68.63} & \textbf{72.44} & \textbf{91.26} & \textbf{80.77} \\
        \midrule 
        Few-Shot+CoT & 1 & 169   & 24    & 30    & 18    & 80.08 & 84.92 & 90.37 & 87.56 \\
        Few-Shot+CoT & 2 & 171   & 23    & 29    & 18    & 80.50 & 85.50 & 90.48 & 87.92 \\
        Few-Shot+CoT & 3 & 171   & 24    & 31    & 15    & 80.91 & 84.65 & 91.94 & 88.14 \\
        Few-Shot+CoT & 4 & 172   & 23    & 29    & 17    & 80.91 & 85.57 & 91.01 & 88.21 \\
        Few-Shot+CoT & 5 & 174   & 22    & 31    & 14    & 81.33 & 84.88 & 92.55 & 88.55 \\
        \textbf{Few-Shot+CoT} & \textbf{Avg} & \textbf{171.4} & \textbf{23.2} & \textbf{30} & \textbf{16.4} & \textbf{80.75} & \textbf{85.11} & \textbf{91.27} & \textbf{88.08} \\     
        \midrule 
        FS+CoT+RAG\_100\% & 1 & 118   & 42    & 25    & 56    & 66.39 & 82.52 & 67.82 & 74.45 \\
        FS+CoT+RAG\_100\% & 2 & 124   & 41    & 26    & 50    & 68.46 & 82.67 & 71.26 & 76.54 \\
        FS+CoT+RAG\_100\% & 3 & 126   & 43    & 23    & 49    & 70.12 & 84.56 & 72.00 & 77.78 \\
        FS+CoT+RAG\_100\% & 4 & 123   & 40    & 27    & 51    & 67.63 & 82.00 & 70.69 & 75.93 \\
        FS+CoT+RAG\_100\%& 5 & 125   & 44    & 24    & 48    & 70.12 & 83.89 & 72.25 & 77.64 \\
        \textbf{FS+CoT+RAG\_100\% } & \textbf{Avg} & \textbf{123.2} & \textbf{42} & \textbf{25} & \textbf{50.8} & \textbf{68.55} & \textbf{83.13} & \textbf{70.80} & \textbf{76.47} \\
        \midrule 
        FS+CoT+RAG\_75\% & 1 & 134   & 50    & 17    & 40    & 76.35 & 88.74 & 77.01 & 82.46 \\
        FS+CoT+RAG\_75\% & 2 & 121   & 45    & 22    & 53    & 68.88 & 84.62 & 69.54 & 76.34 \\
        FS+CoT+RAG\_75\% & 3& 139   & 43    & 24    & 35    & 75.52 & 85.28 & 79.89 & 82.49 \\
        FS+CoT+RAG\_75\% & 4 & 132   & 46    & 21    & 42    & 73.86 & 86.27 & 75.86 & 80.73 \\
        FS+CoT+RAG\_75\% & 5 & 121   & 47    & 20    & 53    & 69.71 & 85.82 & 69.54 & 76.83 \\
        \textbf{FS+CoT+RAG\_75\%} & \textbf{Avg} & \textbf{129.4} & \textbf{46.2} & \textbf{20.8} & \textbf{44.6} & \textbf{72.86} & \textbf{86.14} & \textbf{74.37} & \textbf{79.77} \\
        \midrule 
        FS+CoT+RAG\_50\% & 1 & 115   & 41    & 26    & 59    & 64.73 & 81.56 & 66.09 & 73.02 \\
        FS+CoT+RAG\_50\% & 2 & 132   & 47    & 20    & 42    & 74.27 & 86.84 & 75.86 & 80.98 \\
        FS+CoT+RAG\_50\% & 3 & 120   & 34    & 33    & 54    & 63.90 & 78.43 & 68.97 & 73.39 \\
        FS+CoT+RAG\_50\% & 4 & 126   & 35    & 27    & 53    & 66.80 & 82.35 & 70.39 & 75.90 \\
        FS+CoT+RAG\_50\% & 5 & 123   & 39    & 28    & 51    & 67.22 & 81.46 & 70.69 & 75.69 \\
        \textbf{FS+CoT+RAG\_50\%} & \textbf{Avg} & \textbf{123.2} & \textbf{39.2} & \textbf{26.8} & \textbf{51.8} & \textbf{67.39} & \textbf{82.13} & \textbf{70.40} & \textbf{75.80} \\
        \midrule 
        FS+CoT+RAG\_25\% & 1 & 120   & 48    & 19    & 54    & 69.71 & 86.33 & 68.97 & 76.68 \\
        FS+CoT+RAG\_25\% & 2 & 132   & 41    & 26    & 42    & 71.78 & 83.54 & 75.86 & 79.52 \\
        FS+CoT+RAG\_25\% & 3 & 136   & 43    & 24    & 38    & 74.27 & 85.00 & 78.16 & 81.44 \\
        FS+CoT+RAG\_25\% & 4 & 130   & 46    & 21    & 44    & 73.03 & 86.09 & 74.71 & 80.00 \\
        FS+CoT+RAG\_25\% & 5 & 127   & 45    & 22    & 47    & 71.37 & 85.23 & 72.99 & 78.64 \\
        \textbf{FS+CoT+RAG\_25\%} & \textbf{Avg} & \textbf{129} & \textbf{44.6} & \textbf{22.4} & \textbf{45} & \textbf{72.03} & \textbf{85.24} & \textbf{74.14} & \textbf{79.25} \\
        \midrule 
        FS+CoT+RAG\_0\%  & 1 & 119   & 38    & 36    & 48    & 65.15 & 76.77 & 71.26 & 73.91 \\
        FS+CoT+RAG\_0\%  & 2 & 124   & 42    & 25    & 50    & 68.88 & 83.22 & 71.26 & 76.78 \\
        FS+CoT+RAG\_0\%  & 3 & 131   & 38    & 26    & 46    & 70.12 & 83.44 & 74.01 & 78.44 \\
        FS+CoT+RAG\_0\%  & 4 & 135   & 44    & 22    & 40    & 74.27 & 85.99 & 77.14 & 81.33 \\
        FS+CoT+RAG\_0\%  & 5 & 128   & 39    & 21    & 53    & 69.29 & 85.91 & 70.72 & 77.58 \\
        \textbf{FS+CoT+RAG\_0\% } & \textbf{Avg} & \textbf{127.4} & \textbf{40.2} & \textbf{26} & \textbf{47.4} & \textbf{69.54} & \textbf{83.07} & \textbf{72.88} & \textbf{77.61} \\
    \end{xltabular}

\section{Experimental Trial Results for Code Llama}


{\footnotesize
\setlength{\tabcolsep}{0.2pt}

\sisetup{
  detect-weight=true,
  detect-inline-weight=math,
  table-number-alignment=center
}

\begin{xltabular}{\textwidth}{l|S S S S S |S S S S}
\caption{Experimental Trial Results Using Code Llama}
\label{tab:appendix_c_codellamatrials}\\

\toprule
\textbf{Study} & \textbf{Trial} & \textbf{TP} & \textbf{TN} & \textbf{FP} & \textbf{FN}
& \textbf{Acc(\%)} & \textbf{Prec(\%)} & \textbf{Rec(\%)} & \textbf{F1(\%)} \\
\midrule
\endfirsthead

\toprule
\textbf{Study} & \textbf{Trial} & \textbf{TP} & \textbf{TN} & \textbf{FP} & \textbf{FN}
& \textbf{Acc(\%)} & \textbf{Prec(\%)} & \textbf{Rec(\%)} & \textbf{F1(\%)} \\
\midrule
\endhead

\midrule
\multicolumn{10}{r}{\textit{Continued on next page}}\\
\endfoot

\bottomrule
\endlastfoot
Zero-Shot & 1& 53    & 41    & 26    & 121   & 39.00 & 67.09 & 30.46 & 41.90 \\
Zero-Shot & 2 & 55    & 40    & 27    & 119   & 39.42 & 67.07 & 31.61 & 42.97 \\
Zero-Shot & 3 & 60    & 43    & 24    & 114   & 42.74 & 71.43 & 34.48 & 46.51 \\
Zero-Shot & 4 & 58    & 54    & 13    & 116   & 46.47 & 81.69 & 33.33 & 47.35 \\
Zero-Shot & 5 & 67    & 46    & 21    & 107   & 46.89 & 76.14 & 38.51 & 51.15 \\
\textbf{Zero-Shot} & \textbf{Avg} & \textbf{58.6} & \textbf{44.8} & \textbf{22.2} & \textbf{115.4} & \textbf{42.90} & \textbf{72.68} & \textbf{33.68} & \textbf{45.97} \\
\midrule 
Few-Shot & 1 & 84    & 38    & 29    & 90    & 50.62 & 74.34 & 48.28 & 58.54 \\
Few-Shot & 2 & 75    & 40    & 27    & 99    & 47.72 & 73.53 & 43.10 & 54.35 \\
Few-Shot & 3 & 79    & 37    & 30    & 95    & 48.13 & 72.48 & 45.40 & 55.83 \\
Few-Shot & 4 & 75    & 41    & 26    & 99    & 48.13 & 74.26 & 43.10 & 54.55 \\
Few-Shot & 5 & 86    & 36    & 31    & 88    & 50.62 & 73.50 & 49.43 & 59.11 \\
\textbf{Few-Shot} & \textbf{Avg} & \textbf{79.8} & \textbf{38.4} & \textbf{28.6} & \textbf{94.2} & \textbf{49.05} & \textbf{73.62} & \textbf{45.86} & \textbf{56.47} \\
\midrule 
Few-Shot+CoT & 1& 116   & 35    & 34    & 56    & 62.66 & 77.33 & 67.44 & 72.05 \\
Few-Shot+CoT & 2 & 117   & 39    & 30    & 55    & 64.73 & 79.59 & 68.02 & 73.35 \\
Few-Shot+CoT & 3 & 119   & 35    & 34    & 53    & 63.90 & 77.78 & 69.19 & 73.23 \\
Few-Shot+CoT & 4 & 118   & 37    & 35    & 51    & 64.32 & 77.12 & 69.82 & 73.29 \\
Few-Shot+CoT & 5 & 121   & 38    & 34    & 48    & 65.98 & 78.06 & 71.60 & 74.69 \\
\textbf{Few-Shot+CoT} & \textbf{Avg} & \textbf{118.2} & \textbf{36.8} & \textbf{33.4} & \textbf{52.6} & \textbf{64.32} & \textbf{77.98} & \textbf{69.21} & \textbf{73.32} \\
\midrule 
FS+CoT+RAG\_100\% & 1 & 108   & 35    & 32    & 66    & 59.34 & 77.14 & 62.07 & 68.79 \\
FS+CoT+RAG\_100\% & 2 & 101   & 34    & 33    & 73    & 56.02 & 75.37 & 58.05 & 65.58 \\
FS+CoT+RAG\_100\%& 3& 116   & 34    & 33    & 58    & 62.24 & 77.85 & 66.67 & 71.83 \\
FS+CoT+RAG\_100\% & 4 & 98    & 39    & 28    & 76    & 56.85 & 77.78 & 56.32 & 65.33 \\
FS+CoT+RAG\_100\% & 5 & 104   & 38    & 29    & 70    & 58.92 & 78.20 & 59.77 & 67.75 \\
\textbf{FS+CoT+RAG\_100\%} & \textbf{Avg} & \textbf{105.4} & \textbf{36} & \textbf{31} & \textbf{68.6} & \textbf{58.67} & \textbf{77.27} & \textbf{60.57} & \textbf{67.86} \\
\midrule 
FS+CoT+RAG\_75\% & 1 & 100   & 37    & 30    & 74    & 56.85 & 76.92 & 57.47 & 65.79 \\
FS+CoT+RAG\_75\% & 2 & 106   & 36    & 31    & 68    & 58.92 & 77.37 & 60.92 & 68.17 \\
FS+CoT+RAG\_75\% & 3 & 113   & 36    & 31    & 61    & 61.83 & 78.47 & 64.94 & 71.07 \\
FS+CoT+RAG\_75\% & 4 & 103   & 37    & 30    & 71    & 58.09 & 77.44 & 59.20 & 67.10 \\
FS+CoT+RAG\_75\% & 5 & 96    & 41    & 26    & 78    & 56.85 & 78.69 & 55.17 & 64.86 \\
\textbf{FS+CoT+RAG\_75\%} & \textbf{Avg} & \textbf{103.6} & \textbf{37.4} & \textbf{29.6} & \textbf{70.4} & \textbf{58.51} & \textbf{77.78} & \textbf{59.54} & \textbf{67.40} \\
\midrule 
FS+CoT+RAG\_50\% & 1 & 96    & 36    & 31    & 78    & 54.77 & 75.59 & 55.17 & 63.79 \\
FS+CoT+RAG\_50\% & 2& 99    & 33    & 34    & 75    & 54.77 & 74.44 & 56.90 & 64.50 \\
FS+CoT+RAG\_50\% & 3 & 101   & 35    & 32    & 73    & 56.43 & 75.94 & 58.05 & 65.80 \\
FS+CoT+RAG\_50\% & 4 & 89    & 38    & 29    & 85    & 52.70 & 75.42 & 51.15 & 60.96 \\
FS+CoT+RAG\_50\% & 5 & 103   & 41    & 27    & 70    & 59.75 & 79.23 & 59.54 & 67.99 \\
\textbf{FS+CoT+RAG\_50\%} & \textbf{Avg} & \textbf{97.6} & \textbf{36.6} & \textbf{30.6} & \textbf{76.2} & \textbf{55.68} & \textbf{76.12} & \textbf{56.16} & \textbf{64.61} \\
\midrule
FS+CoT+RAG\_25\%& 1 & 97    & 30    & 46    & 68    & 52.70 & 67.83 & 58.79 & 62.99 \\
FS+CoT+RAG\_25\% & 2 & 107   & 37    & 28    & 69    & 59.75 & 79.26 & 60.80 & 68.81 \\
FS+CoT+RAG\_25\% & 3 & 104   & 34    & 33    & 70    & 57.26 & 75.91 & 59.77 & 66.88 \\
FS+CoT+RAG\_25\% & 4 & 109   & 39    & 28    & 65    & 61.41 & 79.56 & 62.64 & 70.10 \\
FS+CoT+RAG\_25\% & 5 & 93    & 41    & 26    & 81    & 55.60 & 78.15 & 53.45 & 63.48 \\
\textbf{FS+CoT+RAG\_25\%} & \textbf{Avg} & \textbf{102} & \textbf{36.2} & \textbf{32.2} & \textbf{70.6} & \textbf{57.34} & \textbf{76.14} & \textbf{59.09} & \textbf{66.45} \\
\midrule
FS+CoT+RAG\_0\% & 1 & 91    & 37    & 35    & 78    & 53.11 & 72.22 & 53.85 & 61.69 \\
FS+CoT+RAG\_0\%  & 2 & 95    & 31    & 39    & 73    & 52.94 & 70.90 & 56.55 & 62.91 \\
FS+CoT+RAG\_0\%  & 3 & 89    & 42    & 25    & 85    & 54.36 & 78.07 & 51.15 & 61.81 \\
FS+CoT+RAG\_0\%  & 4 & 97    & 40    & 36    & 68    & 56.85 & 72.93 & 58.79 & 65.10 \\
FS+CoT+RAG\_0\%  & 5 & 111   & 42    & 25    & 63    & 63.49 & 81.62 & 63.79 & 71.61 \\
\textbf{FS+CoT+RAG\_0\% } & \textbf{Avg} & \textbf{96.6} & \textbf{38.4} & \textbf{32} & \textbf{73.4} & \textbf{56.15} & \textbf{75.15} & \textbf{56.82} & \textbf{64.63} \\
\end{xltabular}
}

\section{Experimental Trial Results for QWen2.5-coder}
\setcounter{table}{0} 

{\footnotesize
\setlength{\tabcolsep}{0.2pt}

\begin{xltabular}{\textwidth}{l|S S S S S |S S S S}
\caption{Experimental Trial Results Using Qwen2.5-coder}
\label{tab:appendix_d_qwen2.5trials} \\

\toprule
\textbf{Study} & \textbf{Trial} & \textbf{TP} & \textbf{TN} & \textbf{FP} & \textbf{FN} 
& \textbf{Acc (\%)} & \textbf{Prec (\%)} & \textbf{Rec (\%)} & \textbf{F1 (\%)} \\
\midrule
\endfirsthead

\toprule
\textbf{Study} & \textbf{Trial} & \textbf{TP} & \textbf{TN} & \textbf{FP} & \textbf{FN} 
& \textbf{Acc (\%)} & \textbf{Prec (\%)} & \textbf{Rec (\%)} & \textbf{F1 (\%)} \\
\midrule
\endhead

\midrule 
\multicolumn{10}{r}{\textit{Continued on next page}} \\
\endfoot

\bottomrule
\endlastfoot
        Zero-Shot & 1 & 133   & 27    & 40    & 41    & 66.39 & 76.88 & 76.44 & 76.66 \\
        
        Zero-Shot & 2 & 137   & 32    & 35    & 37    & 70.12 & 79.65 & 78.74 & 79.19 \\
        
        Zero-Shot & 3 & 121   & 31    & 36    & 53    & 63.07 & 77.07 & 69.54 & 73.11 \\
        
        Zero-Shot & 4 & 131   & 33    & 34    & 43    & 68.05 & 79.39 & 75.29 & 77.29 \\
        
        Zero-Shot & 5 & 130   & 30    & 37    & 44    & 66.39 & 77.84 & 74.71 & 76.25 \\
        
        \textbf{Zero-Shot} & \textbf{Avg} & \textbf{130.4} & \textbf{30.6} & \textbf{36.4} & \textbf{43.6} & \textbf{66.80} & \textbf{78.17} & \textbf{74.94} & \textbf{76.50} \\
        \hline
        Few-Shot & 1 & 143   & 27    & 40    & 31    & 70.54 & 78.14 & 82.18 & 80.11 \\
        
        Few-Shot & 2 & 147   & 32    & 35    & 27    & 74.27 & 80.77 & 84.48 & 82.58 \\
        
        Few-Shot & 3 & 148   & 36    & 25    & 42    & 73.31 & 85.55 & 77.89 & 81.54 \\
        
        Few-Shot & 4 & 131   & 33    & 34    & 33    & 71.00 & 79.39 & 79.88 & 79.64 \\
        
        Few-Shot & 5 & 140   & 30    & 35    & 39    & 69.67 & 80.00 & 78.21 & 79.10 \\
        
        \textbf{Few-Shot} & \textbf{Avg} & \textbf{141.8} & \textbf{31.6} & \textbf{33.8} & \textbf{34.4} & \textbf{71.76} & \textbf{80.77} & \textbf{80.53} & \textbf{80.59} \\
        \hline
        Few-Shot+CoT & 1 & 153   & 57    & 12    & 19    & 87.14 & 92.73 & 88.95 & 90.80 \\
        
        Few-Shot+CoT & 2 & 156   & 54    & 13    & 18    & 87.14 & 92.31 & 89.66 & 90.96 \\
        
        Few-Shot+CoT & 3 & 154   & 55    & 12    & 20    & 86.72 & 92.77 & 88.51 & 90.59 \\
        
        Few-Shot+CoT & 4 & 155   & 54    & 12    & 20    & 86.72 & 92.81 & 88.57 & 90.64 \\
        
        Few-Shot+CoT & 5 & 156   & 55    & 11    & 19    & 87.55 & 93.41 & 89.14 & 91.23 \\
        
        \textbf{Few-Shot+CoT} & \textbf{Avg} & \textbf{154.8} & \textbf{55} & \textbf{12} & \textbf{19.2} & \textbf{87.05} & \textbf{92.81} & \textbf{88.97} & \textbf{90.84} \\
        \hline
        FS+CoT+RAG\_100\% & 1 & 133   & 33    & 37    & 38    & 68.88 & 78.24 & 77.78 & 78.01 \\
        
        FS+CoT+RAG\_100\% & 2 & 134   & 35    & 35    & 37    & 70.12 & 79.29 & 78.36 & 78.82 \\
        
        FS+CoT+RAG\_100\% & 3 & 128   & 37    & 34    & 42    & 68.46 & 79.01 & 75.29 & 77.11 \\
        
        FS+CoT+RAG\_100\% & 4 & 133   & 32    & 35    & 41    & 68.46 & 79.17 & 76.44 & 77.78 \\
        
        FS+CoT+RAG\_100\% & 5 & 137   & 37    & 28    & 39    & 72.20 & 83.03 & 77.84 & 80.35 \\
        
        \textbf{FS+CoT+RAG\_100\%} & \textbf{Avg} & \textbf{133} & \textbf{34.8} & \textbf{33.8} & \textbf{39.4} & \textbf{69.63} & \textbf{79.75} & \textbf{77.14} & \textbf{78.41} \\
        \hline
        FS+CoT+RAG\_75\% & 1 & 139   & 36    & 36    & 30    & 72.61 & 79.43 & 82.25 & 80.81 \\
        
        FS+CoT+RAG\_75\% & 2 & 136   & 35    & 35    & 35    & 70.95 & 79.53 & 79.53 & 79.53 \\
        
        FS+CoT+RAG\_75\% & 3 & 132   & 29    & 43    & 37    & 66.80 & 75.43 & 78.11 & 76.74 \\
        
        FS+CoT+RAG\_75\% & 4 & 121   & 37    & 35    & 48    & 65.56 & 77.56 & 71.60 & 74.46 \\
        
        FS+CoT+RAG\_75\% & 5 & 130   & 36    & 38    & 37    & 68.88 & 77.38 & 77.84 & 77.61 \\
        
        \textbf{FS+CoT+RAG\_75\%} & \textbf{Avg} & \textbf{131.6} & \textbf{34.6} & \textbf{37.4} & \textbf{37.4} & \textbf{68.96} & \textbf{77.87} & \textbf{77.87} & \textbf{77.83} \\
        \hline
        FS+CoT+RAG\_50\% & 1 & 115   & 38    & 36    & 52    & 63.49 & 76.16 & 68.86 & 72.33 \\
        
        FS+CoT+RAG\_50\% & 2 & 116   & 44    & 37    & 44    & 66.39 & 75.82 & 72.50 & 74.12 \\
        
        FS+CoT+RAG\_50\% & 3 & 114   & 42    & 40    & 45    & 64.73 & 74.03 & 71.70 & 72.84 \\
        
        FS+CoT+RAG\_50\% & 4 & 121   & 41    & 28    & 51    & 67.22 & 81.21 & 70.35 & 75.39 \\
        
        FS+CoT+RAG\_50\% & 5 & 119   & 42    & 36    & 44    & 66.80 & 76.77 & 73.01 & 74.84 \\
        
        \textbf{FS+CoT+RAG\_50\%} & \textbf{Avg} & \textbf{117} & \textbf{41.4} & \textbf{35.4} & \textbf{47.2} & \textbf{65.73} & \textbf{76.80} & \textbf{71.28} & \textbf{73.90} \\
        \hline
        FS+CoT+RAG\_25\% & 1 & 119   & 32    & 36    & 54    & 62.66 & 76.77 & 68.79 & 72.56 \\
        
        FS+CoT+RAG\_25\% & 2 & 129   & 31    & 45    & 36    & 66.39 & 74.14 & 78.18 & 76.11 \\
        
        FS+CoT+RAG\_25\% & 3 & 134   & 30    & 46    & 31    & 68.05 & 74.44 & 81.21 & 77.68 \\
        
        FS+CoT+RAG\_25\% & 4 & 122   & 37    & 35    & 46    & 66.25 & 77.71 & 72.62 & 75.08 \\
        
        FS+CoT+RAG\_25\% & 5 & 121   & 39    & 29    & 52    & 66.39 & 80.67 & 69.94 & 74.92 \\
        
        \textbf{FS+CoT+RAG\_25\%} & \textbf{Avg} & \textbf{125} & \textbf{33.8} & \textbf{38.2} & \textbf{43.8} & \textbf{65.95} & \textbf{76.75} & \textbf{74.15} & \textbf{75.27} \\
        \hline
        FS+CoT+RAG\_0\% & 1 & 115   & 38    & 25    & 63    & 63.49 & 82.14 & 64.61 & 72.33 \\
        
        FS+CoT+RAG\_0\% & 2 & 120   & 35    & 34    & 52    & 64.32 & 77.92 & 69.77 & 73.62 \\
        
        FS+CoT+RAG\_0\% & 3 & 122   & 43    & 27    & 49    & 68.46 & 81.88 & 71.35 & 76.25 \\
        
        FS+CoT+RAG\_0\% & 4 & 123   & 39    & 28    & 51    & 67.22 & 81.46 & 70.69 & 75.69 \\
        
        FS+CoT+RAG\_0\% & 5 & 126   & 41    & 31    & 43    & 69.29 & 80.25 & 74.56 & 77.30 \\
                
        \textbf{FS+CoT+RAG\_0\%} & \textbf{Avg} & \textbf{121.2} & \textbf{39.2} & \textbf{29} & \textbf{51.6} & \textbf{66.56} & \textbf{80.73} & \textbf{70.19} & \textbf{75.04} \\
    \end{xltabular}
    }


\section{Sample code}

\begin{lstlisting}[caption={ChatGPT Failed to identify ``Resource Leak Defect'' caused  by missing call to ``MPI\_Type\_free'' before ``MPI\_Finalize''.},captionpos=b,label={listing:ResourceLeakListing}]
  MPI_Datatype type[size];
  MPI_Type_contiguous(2, MPI_DOUBLE, &type[j]);
   /* Resource Leak Caused By MISSING: MPI_Type_free(&type[j]); */
  MPI_Finalize();
\end{lstlisting}

\begin{lstlisting}[caption={Example of ``Resource Leak`` bug in our proposed bug reference dataset. We combined defective code with its respective defect description(s) explicitly.},captionpos=b,label={lst:ResourceLeakBugReference}]
<EXAMPLE>
#include <mpi.h>
#include <stdio.h>

#define ITERATIONS 1000
#define PARAM_PER_ITERATION 3
#define PARAM_LOST_PER_ITERATION 1

static void myOp(int *invec, int *outvec, int *len, MPI_Datatype *dtype) {
  for (int i = 0; i < *len; i++)
    outvec[i] += invec[i];
}

int main(int argc, char **argv) {
  int nprocs=-1, rank = -1;
  int i=1, stag = 0, rtag = 0;
  int buff_size=1, j=0, size=1;
  MPI_Op op[size]; req1=MPI_REQUEST_NULL;
  int dest = (rank == nprocs - 1) ? (0) : (rank + 1);
  int src = (rank == 0) ? (nprocs - 1) : (rank - 1);
  MPI_Init(&argc, &argv);
  MPI_Comm_size(MPI_COMM_WORLD, &nprocs);
  MPI_Comm_rank(MPI_COMM_WORLD, &rank);
  MPI_Send_init(&buf1, buff_size, type, dest, stag, newcom, &req1);
  MPI_Comm com[size];
  MPI_Comm_dup(MPI_COMM_WORLD, &com[j]);
  MPI_Op_create((MPI_User_function *)myOp, 0, &op[j]); 
  MPI_Finalize();
  printf("Rank %d finished normally\n", rank);
  return 0;
}
</EXAMPLE>
<@\textcolor{red}{<OUTPUT>}@>
<@\textcolor{red}{MPI Resource Leak MPI\_Send\_init(\&buf1, buff\_size, type, dest, stag, newcom, \&req1); on line 30 is missing a matching if(req1 != MPI\_REQUEST\_NULL) MPI\_Request\_free(\&req1);.}@>
<@\textcolor{red}{MPI\_Comm\_dup(MPI\_COMM\_WORLD, \&com[j]); on line 33 is missing a matching MPI\_Comm\_free(\&com[j]);.}@>
<@\textcolor{red}{MPI\_Op\_create((MPI\_User\_function *)myOp, 0, \&op[j]); on line 35 is missing a matching MPI\_Op\_free(\&op[j]);}@>
<@\textcolor{red}{VERDICT: MAJOR ERRORS DETECTED}@>
<@\textcolor{red}{</OUTPUT>}@>
\end{lstlisting}

\begin{lstlisting}[caption={Code Snippet illustrating example of Mismatched parameter tag defect. stag and rtag values are different. MPI\_Bsend stag of 0 never received by MPI\_IRecv with rtag of 1, causing MPI\_Wait to wait indefinitely.},captionpos=b,label={lst:MistatchedTagDefect}]
/* This MPI Program Illustrates a deadlock caused by mismatched communication tag parameter values between ranks. Rank 0 sends a message using MPI_Bsend and a tag value of 0. Rank 1 waits for a message with a tag value of 1. Any other ranks finish normally. Rank 1 blocks indefinitely (deadlock) waiting to receive a message with tag value of one which never arrives. The parameter types match (int). */

  MPI_Comm newcom = MPI_COMM_WORLD;
  MPI_Datatype type = MPI_INT;

  stag=0; rtag=1;/* Parameter Tag Values are different */
  int buf1=rank;
  int buffer_attached_size1 = MPI_BSEND_OVERHEAD + sizeof(int);
  char* buffer_attached1 = (char*)malloc(buffer_attached_size1);
  MPI_Buffer_attach(buffer_attached1, buffer_attached_size1);
  int buf2=-1; MPI_Request req2=MPI_REQUEST_NULL;
  if (rank == 0) {
    MPI_Bsend(&buf1, buff_size, type, dest, stag, newcom); 
    
  } else if (rank == 1) {
    MPI_Irecv(&buf2, buff_size, type, src, rtag, newcom, &req2); 
    MPI_Wait(&req2, MPI_STATUS_IGNORE); 
    /* Indefinite Wait. Rank 1 never completes */
  }
  MPI_Buffer_detach(&buffer_attached1, &buffer_attached_size1);
  free(buffer_attached1);
  if(req2 != MPI_REQUEST_NULL) MPI_Request_free(&req2);
\end{lstlisting}

\end{document}